# $N{=}2$ Structures on Solvable Lie Algebras: the $c{=}9$ Classification

José M Figueroa-O'Farrill[†]

*Department of Physics, Queen Mary and Westfield College*
*Mile End Road, London E1 4NS, UK*

ABSTRACT

Let $\mathfrak{g}$ be a finite-dimensional Lie algebra (not necessarily semisimple). It is known that if $\mathfrak{g}$ is self-dual (that is, if it possesses an invariant metric) then it admits an $N{=}1$ (affine) Sugawara construction. Under certain additional hypotheses, this $N{=}1$ structure admits an $N{=}2$ extension. If this is the case, $\mathfrak{g}$ is said to possess an $N{=}2$ structure. It is also known that an $N{=}2$ structure on a self-dual Lie algebra $\mathfrak{g}$ is equivalent to a vector space decomposition $\mathfrak{g} = \mathfrak{g}_+ \oplus \mathfrak{g}_-$ where $\mathfrak{g}_\pm$ are isotropic Lie subalgebras. In other words, $N{=}2$ structures on $\mathfrak{g}$ are in one-to-one correspondence with Manin triples $(\mathfrak{g}, \mathfrak{g}_+, \mathfrak{g}_-)$. In this paper we exploit this correspondence to obtain a classification of the $c{=}9$ $N{=}2$ structures on solvable Lie algebras. In the process we also give some simple proofs for a variety of Lie algebraic results concerning self-dual Lie algebras admitting symplectic or Kähler structures.

[†] e-mail: J.M.Figueroa@qmw.ac.uk

## §1 Introduction

Finite-dimensional reductive—*i.e.*, direct products of simple and abelian—Lie algebras lie at the heart of many constructions in conformal field theory, string theory, topological field theory, and two-dimensional quantum gravity. Many of the properties of these theories are governed in part by the existence of more complicated underlying algebraic structures which have come to be known loosely as chiral algebras, W-algebras, or vertex operator algebras (VOAs). It is widely believed that the classification of these algebraic structures would take us a good deal closer to a full understanding of the above mentioned physical theories; but alas this proves to be a difficult problem. Nevertheless, the available results suggest that reductive Lie algebras will play an important organizational role in this classification. Indeed one finds a variety of functorial constructions starting from (the loop algebra of) a reductive Lie algebra and resulting in one of these more complicated algebraic structures. The best known construction of this kind is perhaps the (affine) Sugawara construction [1]; but in fact, there are many more: the coset construction [2], the generalized Sugawara construction [3], the (generalized) Drinfel'd–Sokolov reduction [4], as well as their supersymmetric extensions: especially $N{=}1$ [5] [6] [7], and $N{=}2$ [8] [9].

For the (generalized) Drinfel'd–Sokolov reduction one can actually prove [10] that under some mild restrictions on the kind of W-algebras one wants to construct, one does not gain anything by considering nonreductive Lie algebras. But for the other constructions, reductivity is an unnecessary restriction; and one which moreover misses some physically interesting examples. In fact, prompted by the construction in [11], Mohammedi [12] (but see also [13]) showed that the necessary and sufficient condition for a Lie algebra to admit a Sugawara construction (or indeed an $N{=}1$ Sugawara construction) is that it should possess an invariant metric. We call such Lie algebras self-dual[1]. The Sugawara construction based on self-dual Lie algebras had already appeared in the mathematical literature in the work of Lian [14] on the classification of simple VOAs.

Given a self-dual Lie algebra it is natural to investigate the conditions under which there exists an $N{=}2$ extension to the $N{=}1$ Sugawara construction. In the language of $\sigma$-models this is equivalent to the question of which $N{=}1$ supersymmetric WZW models have in addition an $N{=}2$ chiral superconformal symmetry; that is, $(2,0)$ supersymmetry. The necessary and sufficient conditions for an $N$=1 supersymmetric $\sigma$-model to posses $(2,0)$ supersymmetry were worked out in [**15**] by Hull and Witten[2]. Specializing their results to the case of a WZW model, we find that we must impose some extra conditions on the Lie algebra. Indeed, as we will review below from a purely algebraic perspective, the extra conditions are the existence of an integrable complex structure compatible with the metric. It had been shown in [**16**] that any even-dimensional semisimple Lie algebra admits such a structure; but as evinced by the examples in [**17**] and [**18**], there are nonabelian nonsemisimple self-dual Lie algebras admitting such constructions.

---

[1] There does not seem to be a universally adopted nomenclature for Lie algebras with an invariant metric. In the literature they are known both as orthogonal Lie algebras or as self-dual Lie algebras. To avoid confusion with the Lie algebras of the orthogonal groups, we prefer to call them self-dual.

In this paper we further this study and in particular we start the classification of $N$=2 structures with $c$=9, our choice of value of the central charge stemming from the fact that $N$=2 theories with $c$=9 serve as algebraic models for compactification spaces in superstring theory. The Levi-Malcev theorem states that any Lie algebra is a semidirect product of a semisimple Lie algebra and its solvable radical. In this paper we make a first step in the problem by classifying the solvable case. In order to achieve this classification we exploit the correpondence between Manin triples and $N$=2 constructions. This correspondence, implicit in the work of [**16**], was discovered by Parkhomenko [**19**] and further elaborated by Getzler [**20**]. Although these authors only treat reductive Lie algebras, it is clear from their work that this restriction is unnecessary. Although the motivation for this work is rooted in superstring compatification, it is clear that the $N$=2 structures on the (nonabelian) solvable Lie algebras which we classify correspond to noncompact target spaces. However it may be of use to have the full moduli space of $N$=2 structures.

This paper is organized as follows. In Section 2 we review the conditions imposed on a Lie algebra $\mathfrak{g}$ for the existence of an $N$=1 Sugawara construction and for this construction to admit an $N$=2 extension. We will see that the conditions for the former are that $\mathfrak{g}$ be self-dual, and for the latter that it should possess, in addition, a compatible complex structure with vanishing Nijenhuis tensor. Notice that the complex structure need not be invariant. In fact, if it were invariant then the Nijenhuis tensor would vanish identically, but it would also force $\mathfrak{g}$ to be abelian. This follows from the stronger result that a self-dual Lie algebra with a compatible Kähler structure is necessarily abelian.

In Section 3 we make explicit the correspondence with Manin triples. If $\mathfrak{g}$ is any self-dual real Lie algebra with an integrable compatible complex structure, then its complexification $\mathfrak{g}_\mathbb{C}$ is a quasi-triangular Lie bialgebra and $(\mathfrak{g}_\mathbb{C}, \mathfrak{g}_+, \mathfrak{g}_-)$ is a Manin triple, where $\mathfrak{g}_\pm$ are the "(anti)holomorphic" subalgebras. Conversely any complex Manin triple is shown to give rise to an $N$=2 structure on a complex Lie algebra. Generic complex Lie algebras (unlike the semisimple ones) have no real forms, so it may be that an $N$=2 structure on a complex Lie algebra does not come from complexifying an $N$=2 structure on a real Lie algebra. We therefore set out to classify the $N$=2 structures with $c$=9 on complex solvable Lie algebras. This correspondence, together with some of the result of [**13**], implies that classifying the $c$=9 models coming from complex solvable Lie algebras is tantamount to classifying all Manin triples $(\mathfrak{m}, \mathfrak{m}_+, \mathfrak{m}_-)$ where $\mathfrak{m}$ is a complex solvable six-dimensional Lie algebra. Their classification will take up the next three sections.

In Section 4 we classify the three-dimensional solvable complex Lie algebras. The classification of the real three-dimensional Lie algebras is of course classic and due to Bianchi; here we simply extend the base field and organize the information in a way that will be useful later on. In Section 5 we classify the solvable and cosolvable three-dimensional complex Lie bialgebras. We use the relation between Lie bialgebras and matched pairs of Lie algebras. In Section 6 we determine the isomorphism classes of such matched pairs and we discard those solutions which are not solvable.

Finally in Section 7 we exhibit the structure of the solutions we found as self-dual Lie algebras in term of double-extensions, using the structure theorem of Medina and Revoy [**21**]. The paper concludes in Section 8, which summarizes the main results of the paper and includes a brief discussion of reality.

Before we proceed, we should say a word about notation. In this paper we are forced to invent names for Lie algebras which to the best of the author's knowledge, have no accepted name. In doing so, we have chosen whenever possible a mnemonic notation: $\mathfrak{a}_d$ denotes the $d$-dimensional abelian Lie algebra, whereas a $d$-dimensional solvable Lie algebra is denoted by $\mathfrak{s}_d$ or $\mathfrak{s}_{d,n}$, where $n = 1, 2, \ldots$, if there are more than one. Some of these algebras have a parameter and we write it in parethesis following the name of the algebra: for instance, $\mathfrak{s}_{3,3}(\mu)$. All Lie algebras will be understood to be complex Lie algebras unless otherwise adorned, for example $\mathfrak{s}_{3,3}^\mathbb{R}(\mu)$.

---

[2] Strictly speaking this assumes that the $N$=2 superconformal symmetry is realized locally. There exist $\sigma$-models possessing $(2,0)$ supersymmetry and which do not satisfy the conditions of [**15**]. The known such models are related to the ones in [**15**] via duality transformations with respect to an isometry which does not leave invariant the complex structure. In this paper we will restrict ourselves to locally realized supersymmetry.

§2 LIE ALGEBRAS ADMITTING AN $N$=2 STRUCTURE

We start with an $N$=1 affine Lie algebra $\widehat{\mathfrak{g}}_{N=1}$ [5] [6] [7], where $\mathfrak{g}$ is a priori *any* finite-dimensional (real or complex, at this point) Lie algebra. We will later see that we shall be forced to demand that $\mathfrak{g}$ be self-dual. Fix a basis $\langle X_i \rangle$ for $\mathfrak{g}$. The bracket in $\widehat{\mathfrak{g}}_{N=1}$ is defined by the following SOPEs:

$$\mathbb{J}_i(Z)\mathbb{J}_j(W) = \frac{g_{ij}}{Z-W} + f_{ij}{}^k \frac{\theta-\varphi}{Z-W}\mathbb{J}_k(W) + \text{reg.} , \qquad (2.1)$$

where $Z=(z,\theta)$ and $W=(w,\varphi)$ are supercoordinates and $Z-W \equiv z-w-\theta\varphi$. Expanding the superfields $\mathbb{J}_i(Z) = \psi_i(z) + \theta I_i(z)$, we can rewrite (2.1) into the more familiar

$$\begin{aligned}
I_i(z)I_j(w) &= \frac{g_{ij}}{(z-w)^2} + \frac{f_{ij}{}^k I_k(w)}{z-w} + \text{reg.} \\
I_i(z)\psi_j(w) &= \frac{f_{ij}{}^k \psi_k(w)}{z-w} + \text{reg.} \\
\psi_i(z)\psi_j(w) &= \frac{g_{ij}}{z-w} + \text{reg.}
\end{aligned} \qquad (2.2)$$

Associativity of the OPEs demands that the symmetric bilinear defined by $\langle X_i , X_j \rangle = g_{ij}$ be invariant. Under the assumption that it is also nondegenerate, so that $g_{ij}$ is invertible with inverse $g^{ij}$, we can decouple the fermions from the affine currents. Indeed, in terms of the modified currents

$$J_i(z) \equiv I_i(z) - \tfrac{1}{2}g^{j\ell} f_{ij}{}^k (\psi_k \psi_\ell)(z) \qquad (2.3)$$

the OPEs (2.2) become

$$\begin{aligned}
J_i(z)J_j(w) &= \frac{g_{ij} - \tfrac{1}{2}\kappa_{ij}}{(z-w)^2} + \frac{f_{ij}{}^k J_k(w)}{z-w} + \text{reg.} \\
J_i(z)\psi_j(w) &= \text{reg.} \\
\psi_i(z)\psi_j(w) &= \frac{g_{ij}}{z-w} + \text{reg.} ,
\end{aligned} \qquad (2.4)$$

where we have introduced the Killing form $\kappa_{ij} = \operatorname{Tr} \operatorname{ad} X_i \operatorname{ad} X_j = f_{ik}{}^\ell f_{j\ell}{}^k$. This form will in general fail to be nondegenerate.

Let us start with the Sugawara construction for $\widehat{\mathfrak{g}}_{N=1}$ in the form (2.2). We define the superfield

$$\mathbb{T} = G^{ij} D\mathbb{J}_i \mathbb{J}_j + F^{ijk} \mathbb{J}_i \mathbb{J}_j \mathbb{J}_k + Z^i \partial \mathbb{J}_i , \qquad (2.5)$$

and we demand that it should obey an $N$=1 superconformal algebra relative to which $\mathbb{J}_i$ are primary superfields of weight $\tfrac{1}{2}$. In fact, it is sufficient to impose the latter condition, which takes the following form:

$$\mathbb{T}(Z)\mathbb{J}_i = \tfrac{1}{2}\frac{\theta-\varphi}{(Z-W)^2}\mathbb{J}_i + \tfrac{1}{2}\frac{1}{Z-W}D\mathbb{J}_i + \frac{\theta-\varphi}{Z-W}\partial\mathbb{J}_i + \text{reg.} . \qquad (2.6)$$

Expanding the superfield $\mathbb{T}(Z) = \tfrac{1}{2}\mathsf{G}(z) + \theta\mathsf{T}(z)$, we find that (2.6) becomes

$$\begin{aligned}
\mathsf{G}(z)\psi_i(w) &= \frac{I_i(w)}{z-w} + \text{reg.} \\
\mathsf{G}(z)I_i(w) &= \frac{\psi_i(w)}{(z-w)^2} + \frac{\partial\psi_i(w)}{z-w} + \text{reg.} \\
\mathsf{T}(z)\psi_i(w) &= \frac{\tfrac{1}{2}\psi_i(w)}{(z-w)^2} + \frac{\partial\psi_i(w)}{z-w} + \text{reg.} \\
\mathsf{T}(z)I_i(w) &= \frac{I_i(w)}{(z-w)^2} + \frac{\partial I_i(w)}{z-w} + \text{reg.}
\end{aligned} \qquad (2.7)$$

It turns out within the Ansatz (2.5), it is enough to satisfy the first equation in (2.7). Indeed, from this equation we find the following conditions [12] on the parameters in (2.5): $Z^i = 0$, $g_{ij}$ is invertible with inverse $2G^{ij}$ (so that $G^{ij} = \tfrac{1}{2}g^{ij}$), and $F^{ijk} = \tfrac{1}{6}f^{ijk}$, where we use $g^{ij}$ and $g_{ij}$ to raise and lower indices.

Since $g_{ij}$ is invertible, we can modify the currents as in (2.3). In terms of the modified currents, the generators of the $N$=1 superconformal algebra become

$$\begin{aligned}
\mathsf{G} &= g^{ij}J_i\psi_j - \tfrac{1}{6}f^{ijk}\psi_i\psi_j\psi_k \\
\mathsf{T} &= \tfrac{1}{2}g^{ij}J_iJ_j + \tfrac{1}{2}g^{ij}\partial\psi_i\psi_j .
\end{aligned} \qquad (2.8)$$

They obey an $N$=1 superconformal algebra with central charge given by

$$c \equiv c(\mathfrak{g}) = \tfrac{3}{2}\dim\mathfrak{g} - \tfrac{1}{2}g^{ij}\kappa_{ij} . \qquad (2.9)$$

Let us now investigate the existence of an $N$=2 extension to this $N$=1 superconformal algebra. Let $\mathsf{G}_1 \equiv \mathsf{G}$ and let us propose the following Ansatz for the second supersymmetry generator:

$$\mathsf{G}_2 = \omega^{ij}J_i\psi_j + \tfrac{1}{6}s^{ijk}\psi_i\psi_j\psi_k + e^i\partial\psi_i . \qquad (2.10)$$

Demanding that it be a primary with respect to $\mathsf{T}$ we already see that $e^i = 0$.

We now change basis to

$$\mathsf{G}^\pm \equiv \tfrac{1}{\sqrt{2}} \left(\mathsf{G}_1 \pm i\mathsf{G}_2\right) = A_\pm^{ij} J_i \psi_j + B_\pm^{ijk} \psi_i \psi_j \psi_k ~, \qquad (2.11)$$

where

$$\begin{aligned} A_\pm^{ij} &= \tfrac{1}{\sqrt{2}} \left(g^{ij} \pm i\omega^{ij}\right) \\ B_\pm^{ijk} &= -\tfrac{1}{6\sqrt{2}} \left(f^{ijk} \mp i s^{ijk}\right) ~. \end{aligned} \qquad (2.12)$$

In [22] we derived the conditions that $A_\pm^{ij}$ and $B_\pm^{ijk}$ must satisfy for $\mathsf{G}^\pm$ to generate an $N=2$ superconformal algebra. Inserting the expression (2.12) into these equations we find that the following conditions must be satisfied [16] [19] [18]. First of all we have that $\omega^{ij}$ is antisymmetric and that it must satisfy the following condition:

$$\omega^{ik} \omega^{j\ell} g_{k\ell} = g^{ij} ~. \qquad (2.13)$$

To interpret this relation it is convenient to define $J^i{}_j \equiv \omega^{ik} g_{kj}$. Then (2.13) says that the linear map $J: \mathfrak{g} \to \mathfrak{g}$ defined by $JX_i = J^j{}_i X_j$ obeys $J^2 = -1$, whence it is an almost complex structure on $\mathfrak{g}$. The antisymmetry of $\omega^{ij}$ says that $J$ is compatible with the metric

$$\langle JX, JY \rangle = \langle X, Y \rangle \qquad \text{for all } X, Y \in \mathfrak{g} ~. \qquad (2.14)$$

This allows us to define a nondegenerate skewsymmetric bilinear form on $\mathfrak{g}$

$$\omega(X, Y) \equiv \langle X, JY \rangle ~. \qquad (2.15)$$

Evaluating it on the basis we see that $\omega(X_i, X_j) = \omega_{ij} = g_{ik} g_{j\ell} \omega^{k\ell}$.

The final two conditions coming from the equations in [22] are

$$s^{ijk} = J^i{}_\ell J^j{}_m J^k{}_n f^{\ell mn} \qquad (2.16)$$

which defines $s^{ijk}$, and

$$f^{ijk} = -J^j{}_m J^k{}_\ell f^{im\ell} - J^i{}_\ell J^j{}_m f^{m\ell k} + J^i{}_\ell J^k{}_m f^{mj\ell} ~. \qquad (2.17)$$

This last equation may seem formidable, but it is in fact equivalent to the vanishing of the Nijenhuis tensor associated to the complex structure $J$:

$$N(X, Y) \equiv [X, Y] - [JX, JY] + J[X, JY] + J[JX, Y] ~. \qquad (2.18)$$

In summary, provided that $\mathfrak{g}$ has an invariant metric with a compatible complex structure with vanishing Nijenhuis tensor, one has an $N=2$ superconformal algebra extending the $N=1$ Sugawara construction. If this is the case, we shall say that $\mathfrak{g}$ *admits an $N=2$ structure*.

The $N=2$ superconformal algebra is generated by $\mathsf{T}$ and $\mathsf{G}_1 = \mathsf{G}$ given by (2.8), and in addition the two generators:

$$\mathsf{G}_2 = -g^{ij} J_i \widetilde{\psi}_j + \tfrac{1}{6} f^{ijk} \widetilde{\psi}_i \widetilde{\psi}_j \widetilde{\psi}_k ~, \qquad (2.19)$$

where $\widetilde{\psi}_i \equiv J^j{}_i \psi_j$; and $U(1)$ current

$$\begin{aligned} \mathsf{J} &= \tfrac{i}{2} \omega^{jk} f_{jk}{}^i J_i - \tfrac{i}{2} \left(\omega^{ij} + \omega^{k\ell} f_{km}{}^i f_{\ell n}{}^j g^{mn}\right) \psi_i \psi_j \\ &= \tfrac{i}{2} \omega^{jk} f_{jk}{}^i I_i - \tfrac{i}{2} \omega^{ij} \psi_i \psi_j ~. \end{aligned} \qquad (2.20)$$

Moreover the central charge of the $N=2$ superconformal algebra is given by (2.9).

The above conditions have a direct geometrical interpretation. Let $G$ be any Lie group with Lie algebra $\mathfrak{g}$. Under the identification of $\mathfrak{g}$ with the left-invariant vector fields, an almost complex structure on $\mathfrak{g}$ is equivalent to a left-invariant almost complex structure on $G$. Moreover the vanishing of the Nijenhuis tensor implies (by the Newlander–Nirenberg theorem) that the complex structure is integrable, so that $G$ becomes a complex Lie group. Similarly, the invariant metric $\langle -, - \rangle$ on $\mathfrak{g}$ gives rise to a bi-invariant metric on $G$, and (2.14) says that the metric and the complex structures are compatible. Therefore the above $N=2$ construction is associated to the following geometric data: a Lie group admitting a bi-invariant metric and a compatible left-invariant (integrable) complex structure.

These geometric conditions are the specialization to the case of a WZW model of the conditions found by Hull and Witten in [15]. They appeared explicitly for the first time by the Leuven group in [16], who considered semisimple Lie groups. The algebraic conditions for self-dual Lie algebras appeared for the first time in work of Mohammedi [18].

Invariant complex structures and Kähler structures

It is natural when adding extra structures on Lie algebras to impose that they be invariant or at least compatible in some sense with the algebraic structure. In this section we analyze what happens when we impose that the complex structure be invariant. We again let $\mathfrak{g}$ be a Lie algebra with an invariant metric and a compatible complex structure. If the complex structure $J : \mathfrak{g} \to \mathfrak{g}$ is invariant:

$$J[X, Y] = [X, JY] \qquad \text{for all } X, Y \in \mathfrak{g} ~. \qquad (2.21)$$

then the Nijenhuis tensor vanishes identically. In fact, one has $[JX, JY] = -[X, Y]$ for all $X, Y \in \mathfrak{g}$. And from this and (2.21) it follows from (2.18) that $N(X, Y) = 0$ for all $X, Y \in \mathfrak{g}$. Therefore any self-dual Lie algebra with an invariant compatible complex structure admits an $N$=2 structure. However we will see below that these algebras are precisely the (even-dimensional) abelian Lie algebras. It is clear that any abelian Lie algebra is self-dual, and that any even-dimensional abelian Lie algebra admits an invariant complex structure. We will see now the converse also holds. In the process we will find other results about self-dual Lie algebras with antisymmetric bilinear forms.

It is first of all clear that a self-dual Lie algebra with a compatible invariant complex structure cannot be semisimple. Indeed, if $\mathfrak{g}$ were semisimple, then we complexify and go to a compact real form. The symplectic form $\omega$ defined by (2.15) is invariant and therefore defines a bi-invariant two-form on the Lie group which is therefore closed. Since the group is compact and symplectic $H^2(G) \neq 0$, but for a semisimple Lie group $H^2(G) = 0$. In fact, this argument says more. A bi-invariant differential form is actually harmonic. By the Hodge decomposition theorem it has to vanish, since it represents the zero class. Thus we see that a bi-invariant two-form in a semisimple Lie algebra is not just necessarily degenerate, but it must in fact vanish.

We can argue algebraically and get a stronger result. Let $\omega$ denote an invariant skewsymmetric bilinear form on $\mathfrak{g}$ and let $\omega_{ij}$ denote its components relative to a fixed basis for $\mathfrak{g}$. Then consider $\alpha^k = \omega^{ij} f_{ij}{}^k$ where we raised the indices of $\omega$ with the invariant metric. Then $\alpha^k$ is invariant since it is constructed out of invariants; in other words, $\alpha^k f_{k\ell}{}^m = 0$. We now massage this equation:

$$\begin{aligned} 0 &= \omega^{ij} f_{ij}{}^k f_{k\ell}{}^m \\ &= -\omega^{ij}\left(f_{j\ell}{}^k f_{ki}{}^m + f_{\ell i}{}^k f_{kj}{}^m\right) && \text{by Jacobi} \\ &= \omega^{mi} f_{j\ell}{}^k f_{ki}{}^j + \omega^{jm} f_{\ell i}{}^k f_{kj}{}^i && \text{by invariance of } \omega \\ &= \omega^{mi} \kappa_{i\ell} - \omega^{jm} \kappa_{\ell j} && \text{by definition of } \kappa \\ &= 2\omega^{mi} \kappa_{i\ell} \ . \end{aligned}$$

Therefore, we see that if $\kappa$ is nondegenerate then $\omega = 0$, which recovers the previous conclusion. But notice that the above computation can also be read as saying that if the symplectic form $\omega$ is nondegenerate, then the Killing form is zero. In other words, we have proven the following

PROPOSITION 2.22. *A Lie algebra with an invariant symplectic form has vanishing Killing form. In particular, it is solvable.* □

These results are actually not as strong as what can be proven if we take into account the vanishing (2.17) of the Nijenhuis tensor associated to the complex structure. In fact, one can prove the following

PROPOSITION 2.23. *Let $\mathfrak{g}$ be self-dual Lie algebra with a compatible integrable complex structure and such that the associated bilinear form* (2.15) *is a cocycle. Then $\mathfrak{g}$ is abelian.*

PROOF: Let $\omega$ be defined by (2.15). We assume that it is a cocycle, so that for all $X, Y, Z \in \mathfrak{g}$,

$$d\omega(X,Y,Z) \equiv \omega([X,Y],Z) + \omega([Y,Z],X) + \omega([Z,X],Y) = 0 \ , \tag{2.24}$$

or relative to the chosen basis

$$f_{ij}{}^\ell \omega_{\ell k} + f_{jk}{}^\ell \omega_{\ell i} + f_{ki}{}^\ell \omega_{\ell i} = 0 \ . \tag{2.25}$$

From (2.15) and using that $N(X,Y) = 0$, we find

$$\begin{aligned} \omega([X,Y],Z) &= \langle [X,Y], JZ\rangle \\ &= \langle [JX,JY], JZ\rangle - \langle J[X,JY], JZ\rangle - \langle J[JX,Y], JZ\rangle \\ &= \langle [JX,JY], JZ\rangle - \langle [X,JY], Z\rangle - \langle [JX,Y], Z\rangle \\ &= \langle [JX,JY], JZ\rangle + \langle JY, [X,Z]\rangle - \langle JX, [Y,Z]\rangle \ , \end{aligned}$$

where in the next to last line we have used the orthogonality of $J$ and in the last line the invariance of the metric. If we now insert this into (2.24), we get

$$\langle [JX, JY], JZ\rangle = 0 \ , \quad \text{for all } X, Y, Z \in \mathfrak{g} \ . \tag{2.26}$$

Since $\langle -, -\rangle$ is nondegenerate and $J$ is invertible, we see that $[JX, JY] = 0$ for all $X, Y \in \mathfrak{g}$. That is, $\mathfrak{g}$ is abelian. □

Notice that this means that a connected Lie group with a bi-invariant metric and a left-invariant Kähler structure is abelian. In particular, any compact Kähler Lie group is a torus. Geometric results of this kind can be found in the mathematical literature. For example, Hano [**23**] (but see also [**24**] and [**25**]) proved that if $G$ is a connected unimodular Lie group with a left-invariant symplectic form, then $G$ is solvable. Hano also proved that any connected nilpotent Lie group with a left-invariant Kähler structure in necessarily abelian.

Finally, we have the following

COROLLARY 2.27. *If $\mathfrak{g}$ is a self-dual Lie algebra with an invariant complex structure, it is abelian.*

PROOF: If $\mathfrak{g}$ has an invariant complex structure $J$, then the Nijenhuis tensor vanishes identically and $\omega$ is invariant. The result follows from Proposition 2.23 after noticing that if $\omega$ is invariant it obeys the cocycle condition (2.24). Indeed, taking (2.15) into account, we find

$$\begin{aligned} d\omega(X,Y,Z) &= \langle [X,Y], JZ\rangle + \langle [Y,Z], JX\rangle + \langle [Z,X], JY\rangle \\ &= \langle X, [Y,JZ]\rangle + \langle Y, [Z,JX]\rangle + \langle Z, [X,JY]\rangle \\ &= \langle X, J[Y,Z]\rangle + \langle Y, J[Z,X]\rangle + \langle Z, J[X,Y]\rangle \quad \text{by (2.21)} \\ &= -\langle JX, [Y,Z]\rangle - \langle JY, [Z,X]\rangle - \langle JZ, [X,Y]\rangle \\ &= -d\omega(X,Y,Z) \ ; \end{aligned}$$

whence $\omega$ is a cocycle. $\square$

§3 The correspondence with Manin triples

The conditions derived in the previous section for the existence of an $N$=2 structure were interpreted by Parkhomenko [19] (and elaborated by Getzler [20]) in terms of Manin triples. Although these papers deal only with reductive Lie algebras, it is clear from their results (in particular from Getzler's) that this restriction is unnecessary. Since we will need this correspondence we review this here briefly and in so doing will derive the same results in a slightly different way. This might shed some further insight into the correspondence.

Let $\mathfrak{g}$ be a real Lie algebra admitting an $N$=2 structure, and let $\mathfrak{g}_{\mathbb{C}}$ be its complexification. We extend the Lie bracket, the invariant metric and the complex structure $\mathbb{C}$-linearly. We break up $\mathfrak{g}_{\mathbb{C}}$ as $\mathfrak{g}_{\mathbb{C}} = \mathfrak{g}_+ \oplus \mathfrak{g}_-$, where $\mathfrak{g}_\pm$ are the $(\pm i)$-eigenspaces of the complex structure $J$. We have the following easy results.

LEMMA 3.1. *The Nijenhuis tensor (2.18) vanishes if and only if $\mathfrak{g}_\pm$ are Lie subalgebras: $[\mathfrak{g}_\pm, \mathfrak{g}_\pm] \subset \mathfrak{g}_\pm$.*

PROOF: Evaluate $N(X,Y)$ for the following cases: $X,Y \in \mathfrak{g}_\pm$ and $X \in \mathfrak{g}_\pm$, $Y \in \mathfrak{g}_\mp$. The lemma follows. $\square$

LEMMA 3.2. *The compatibility condition (2.14) implies that $\mathfrak{g}_\pm$ are isotropic subalgebras.*

PROOF: If $X,Y \in \mathfrak{g}_\pm$ then $\langle X, Y\rangle = -\langle X, Y\rangle$, hence it is zero. $\square$

To interpret this we need to recall the concept of a Lie bialgebra. A Lie bialgebra is a Lie algebra $\mathfrak{b}$ and a cobracket $\mathfrak{b} \to \bigwedge^2 \mathfrak{b}$ whose transpose defines a Lie bracket on the dual $\mathfrak{b}^*$. The cobracket must also satisfy the cocycle condition below. These conditions can be summarized by saying that $(\mathfrak{b} \oplus \mathfrak{b}^*, \mathfrak{b}, \mathfrak{b}^*)$ is a Manin triple (see, for example, [26]). In other words, $\mathfrak{b} \oplus \mathfrak{b}^*$ is a Lie algebra containing $\mathfrak{b}$ and $\mathfrak{b}^*$ as Lie subalgebras and where the bracket $[\mathfrak{b}, \mathfrak{b}^*]$ is fixed by demanding invariance of the natural inner product on $\mathfrak{b} \oplus \mathfrak{b}^*$ induced by the dual pairing between $\mathfrak{b}$ and $\mathfrak{b}^*$.

In other words, Lemma 3.1 and Lemma 3.2 imply that $(\mathfrak{g}_{\mathbb{C}}, \mathfrak{g}_+, \mathfrak{g}_-)$ is a Manin triple [26], whence $\mathfrak{g}_+$, $\mathfrak{g}_-$ and $\mathfrak{g}_{\mathbb{C}}$ are Lie bialgebras. In particular $\mathfrak{g}_{\mathbb{C}}$ is quasi-triangular. Notice that $\mathfrak{g}$ can be recovered from $\mathfrak{g}_{\mathbb{C}}$ as the fixed point set under complex conjugation; that is, as the real form under the antilinear involutive automorphism mapping $\mathfrak{g}_\pm \to \mathfrak{g}_\mp$.

There is a partial converse to this result; namely, to any complex Manin triple there is associated an $N$=2 superconformal algebra. This is asserted in [19] and proved in [20] without making any essential use of the assumptions of reductivity of the Lie algebras. Here we will not repeat Getzler's proof but rather give an alternate proof of this result by simply deriving the form of their $N$=2 generators from the ones obtained in the previous section. We say that this is only a partial converse, because the existence of a real form is a separate issue. We will comment briefly on this later on.

To this end, let us define the following projectors $\mathbb{P}_\pm \equiv \frac{1}{2}(1 \mp iJ) : \mathfrak{g}_{\mathbb{C}} \to \mathfrak{g}_{\mathbb{C}}$. It follows that $\mathbb{P}_\pm$ is the projector onto $\mathfrak{g}_\pm$ along $\mathfrak{g}_\mp$. Lemma 3.2 says that relative to the metric on $\mathfrak{g}_{\mathbb{C}}$, $\mathbb{P}_+$ and $\mathbb{P}_-$ are mutually adjoint.

Let $\langle X_i\rangle$ be a $\mathbb{C}$-basis for $\mathfrak{g}_{\mathbb{C}}$ and let $\mathbf{J} = J^i X_i$ and $\mathbf{\Psi} = \psi^i X_i$ be Lie algebra valued fields, where we raise and lower indices with $g^{ij}$ and $g_{ij}$. We let $\mathbf{J}_\pm = \mathbb{P}_\pm \mathbf{J}$ and $\mathbf{\Psi}_\pm = \mathbb{P}_\pm \mathbf{\Psi}$ denote their projections onto $\mathfrak{g}_\pm$. Consider now the expressions for the supersymmetry generators $\mathsf{G}^\pm$ given by inserting the first equation in (2.8) and (2.19) into (2.11):

$$\mathsf{G}^\pm = \tfrac{1}{\sqrt{2}} g^{ij} J_i \left( \psi_j \mp i\widetilde{\psi}_j \right) - \tfrac{1}{6\sqrt{2}} f^{ijk} \left( \psi_i \psi_j \psi_k \mp i\widetilde{\psi}_i \widetilde{\psi}_j \widetilde{\psi}_k \right) \ , \qquad (3.3)$$

where again $\widetilde{\psi}_i \equiv J^j{}_i \psi_j$. If we define $\widetilde{\mathbf{\Psi}} = \widetilde{\psi}^i X_i$, then we can write the above expressions in a more invariant form:

$$\mathsf{G}^\pm = \tfrac{1}{\sqrt{2}} \langle \mathbf{J}, \mathbf{\Psi} \mp i\widetilde{\mathbf{\Psi}}\rangle - \tfrac{1}{6\sqrt{2}} \left( \langle [\mathbf{\Psi}, \mathbf{\Psi}], \mathbf{\Psi}\rangle \mp i \langle [\widetilde{\mathbf{\Psi}}, \widetilde{\mathbf{\Psi}}], \widetilde{\mathbf{\Psi}}\rangle \right) \ . \qquad (3.4)$$

Now notice the following identities:

$$\mathbf{\Psi} = \mathbf{\Psi}_+ + \mathbf{\Psi}_- \quad \text{and} \quad \widetilde{\mathbf{\Psi}} = i(\mathbf{\Psi}_+ - \mathbf{\Psi}_-) \ . \qquad (3.5)$$

We now insert them into (3.4) and manipulate the resulting expressions. First of all notice that $\mathbf{\Psi} \mp i\widetilde{\mathbf{\Psi}} = 2\mathbf{\Psi}_\pm$, whence the $\mathbf{J}$-dependent term becomes $\sqrt{2}\langle \mathbf{J}, \mathbf{\Psi}_\pm\rangle = \sqrt{2}\langle \mathbf{J}_\mp, \mathbf{\Psi}_\pm\rangle$, where the last equality follows by isotropy of $\mathfrak{g}_\pm$. Inserting (3.5) into the first trilinear term of (3.4) we find

$$\begin{aligned} \langle [\mathbf{\Psi}, \mathbf{\Psi}], \mathbf{\Psi}\rangle &= \langle [\mathbf{\Psi}_+ + \mathbf{\Psi}_-, \mathbf{\Psi}_+ + \mathbf{\Psi}_-], \mathbf{\Psi}_+ + \mathbf{\Psi}_-\rangle \\ &= \langle [\mathbf{\Psi}_+, \mathbf{\Psi}_+], \mathbf{\Psi}_-\rangle + \langle [\mathbf{\Psi}_-, \mathbf{\Psi}_-], \mathbf{\Psi}_+\rangle + 2\langle [\mathbf{\Psi}_-, \mathbf{\Psi}_+], \mathbf{\Psi}_+ + \mathbf{\Psi}_-\rangle \\ &= 3\langle [\mathbf{\Psi}_+, \mathbf{\Psi}_+], \mathbf{\Psi}_-\rangle + 3\langle [\mathbf{\Psi}_-, \mathbf{\Psi}_-], \mathbf{\Psi}_+\rangle \ , \qquad (3.6) \end{aligned}$$

where we repeatedly used the fact that $\mathfrak{g}_\pm$ are isotropic subalgebras and that the metric is invariant. Similarly for the second trilinear term we find

$$\begin{aligned} i\langle[\widetilde{\Psi},\widetilde{\Psi}],\widetilde{\Psi}\rangle &= \langle[\Psi_+ - \Psi_-, \Psi_+ - \Psi_-], \Psi_+ - \Psi_-\rangle \\ &= -\langle[\Psi_+,\Psi_+],\Psi_-\rangle + \langle[\Psi_-,\Psi_-],\Psi_+\rangle - 2\langle[\Psi_-,\Psi_+],\Psi_+ - \Psi_-\rangle \\ &= -3\langle[\Psi_+,\Psi_+],\Psi_-\rangle + 3\langle[\Psi_-,\Psi_-],\Psi_+\rangle \ . \end{aligned} \quad (3.7)$$

Putting it all together we find

$$\mathsf{G}^\pm = \sqrt{2}\langle \mathbf{J}_\mp, \Psi_\pm\rangle - \tfrac{1}{\sqrt{2}}\langle[\Psi_\pm,\Psi_\pm],\Psi_\mp\rangle \ . \quad (3.8)$$

Notice that the above formula does not make reference to the complex structure on $\mathfrak{g}$ (hence $\mathfrak{g}_\mathbb{C}$); instead, it only makes reference to the structures present in a Manin triple. Hence if $(\mathfrak{m},\mathfrak{m}_+,\mathfrak{m}_-)$ is *any* Manin triple, the generators in (3.8) define an $N$=2 superconformal algebra. Moreover, we can recover the complex structure by $J = i(\mathbb{P}_+ - \mathbb{P}_-)$. It is easy to show that $J^2 = -1$, that it is compatible with the metric, and that its Nijenhuis tensor vanishes. Thus from the results of the previous section, the fields in (3.8) indeed generate an $N$=2 superconformal algebra.

Now suppose that in addition there exists an antilinear involutive automorphism $\sigma : \mathfrak{m} \to \mathfrak{m}$ mapping $\mathfrak{m}_\pm \to \mathfrak{m}_\mp$, then the Lie subalgebra $\mathfrak{m}_\mathbb{R} \subset \mathfrak{m}$ fixed by $\sigma$ also admits an invariant metric and a compatible complex structure with vanishing Nijenhuis tensor. The existence of a conjugation $\sigma$, which can be proven for the reductive examples, is not guaranteed in the case of general Lie algebras. Indeed, complex Lie algebras need not have real forms. One can understand this from deformation theory. Contrary to semisimple Lie algebras, generic Lie algebras are not rigid under deformations. Over the complex numbers, the deformation parameter may be taken to be complex, and in general no change of basis can get rid of its imaginary part.

We finally make contact with the expressions for the $N$=2 generators given in [19] and [20]. Let $(\mathfrak{m},\mathfrak{m}_+,\mathfrak{m}_-)$ be a complex Manin triple and let us choose dual bases $\langle e_i\rangle$ and $\langle e^i\rangle$ for $\mathfrak{m}_+$ and $\mathfrak{m}_- \cong \mathfrak{m}_+^*$ respectively. Let the bracket of $\mathfrak{m}_+$ be given by $[e_i,e_j] = f_{ij}{}^k e_k$ and that of $\mathfrak{m}_-$ by $[e^i,e^j] = c^{ij}{}_k e^k$. We write $[e_i,e^j] = A_{ik}{}^j e^k + B^{jk}{}_i e_k$ and we determine $A$ and $B$ uniquely by demanding that the dual pairing $\langle e_i, e^j\rangle = \delta_i^j$ be invariant. Doing so we find

$$[e_i,e^j] = c^{jk}{}_i e_k - f_{ik}{}^j e^k \ . \quad (3.9)$$

Imposing the Jacobi identity for the bracket on $\mathfrak{m}$ we recover the Jacobi identities for $\mathfrak{m}_+$ and for $\mathfrak{m}_-$ and in addition the cocycle condition

$$f_{ij}{}^m c^{k\ell}{}_m = f_{im}{}^k c^{m\ell}{}_j - f_{im}{}^\ell c^{mk}{}_j - f_{jm}{}^k c^{m\ell}{}_i + f_{jm}{}^\ell c^{mk}{}_i \ . \quad (3.10)$$

Relative to this basis, the generators in (3.8) take the form:

$$\begin{aligned} \mathsf{G}^+ &= \sqrt{2}\left(J_i \psi^i - \tfrac{1}{2} f_{ij}{}^k \psi^i \psi^j \psi_k\right) \\ \mathsf{G}^- &= \sqrt{2}\left(J^i \psi_i - \tfrac{1}{2} c^{ij}{}_k \psi_i \psi_j \psi^k\right) \end{aligned} \quad (3.11)$$

where $J_i = \langle \mathbf{J}, e_i\rangle$ and $J^i = \langle \mathbf{J}, e^i\rangle$ and similarly for the fermions. Then these fields indeed obey an $N$=2 superconformal algebra together with

$$\begin{aligned} \mathsf{J} &= \psi^i \psi_i + f_{ji}{}^i I^j + c^{ji}{}_i I_j \\ \mathsf{T} &= J^i J_i + \tfrac{1}{2}\partial\psi_i \psi^i - \tfrac{1}{2}\psi_i \partial\psi^i \ , \end{aligned} \quad (3.12)$$

where the coupled currents $I_i$ and $I^i$ are given by

$$\begin{aligned} I_i &= J_i - f_{ij}{}^k \psi^j \psi_k - \tfrac{1}{2} c^{jk}{}_i \psi_j \psi_k \\ I^i &= J^i - c^{ij}{}_k \psi_j \psi^k - \tfrac{1}{2} f_{jk}{}^i \psi^j \psi^k \ . \end{aligned} \quad (3.13)$$

The Virasoro central charge can be computed from (2.9). In terms of the above basis we find

$$\begin{aligned} c &= 3\dim\mathfrak{m}_+ - \kappa(e^i,e_i) \\ &= 3\dim\mathfrak{m}_+ - \langle e_j, [e^i,[e_i,e^j]]\rangle - \langle e^j, [e^i,[e_i,e_j]]\rangle \\ &= 3\dim\mathfrak{m}_+ + \langle[e^i,e_j],[e_i,e^j]\rangle + \langle[e^i,e^j],[e_i,e_j]\rangle \\ &= 3\dim\mathfrak{m}_+ + 3c^{ij}{}_k f_{ij}{}^k \\ &= 3\dim\mathfrak{m}_+ - 6c^{kj}{}_j f_{ik}{}^i \ , \end{aligned} \quad (3.14)$$

where the identity responsible for the last equality was obtained from the cocycle condition (3.10) by contracting the free indices in the obvious fashion.

In summary, there is a one-to-one correspondence between complex Manin triples and $N$=2 structures on complex Lie algebras. In trying to classify $N$=2 structures on (complex) Lie algebras we will instead classify the corresponding complex Manin triples. We should emphasize that the data involved in an $N$=2 structure is precisely the Lie bracket and the metric of the larger Lie algebra in the Manin triple *as well as* the choice of complementary isotropic subalgebras $\mathfrak{m}_\pm$. Equivalently, the data $(\mathfrak{m},\mathfrak{m}_+,\mathfrak{m}_-)$ of a complex Manin triple corresponds precisely to two Lie algebras $\mathfrak{m}_+$ and $\mathfrak{m}_-$ *and* a nondegenerate pairing $\mathfrak{m}_+ \otimes \mathfrak{m}_- \to \mathbb{C}$, in such a way that the unique Lie brackets on $\mathfrak{m}_+ \oplus \mathfrak{m}_-$ leaving this pairing invariant obey the Jacobi identity. If this is the case one says that $\mathfrak{m}_+$ and $\mathfrak{m}_-$ are *matched*. The rest of the paper is dedicated to the classification of all matched pairs $(\mathfrak{m}_+, \mathfrak{m}_-)$, where all three Lie algebras in the associated Manin triple $(\mathfrak{m}, \mathfrak{m}_+, \mathfrak{m}_-)$ are solvable, and where $\mathfrak{m}_\pm$ are three-dimensional. The condition on the dimension is to ensure that the Virasoro central charge of the $N$=2 structure equals 9, as we will see below. But before continuing with the main thread of the paper we digress briefly to mention

An intriguing connection

Notice that the complex structure $J$ associated to a complex Manin triple $(\mathfrak{m}, \mathfrak{m}_+, \mathfrak{m}_-)$ is given by $J = iR$, where $R = \mathbb{P}_+ - \mathbb{P}_-$. It turns out that $R$ is a classical $r$-matrix and it satisfies the (modified) classical Yang–Baxter equation. Let us elaborate briefly on this.

Suppose that $\mathfrak{g}$ is a Lie algebra and let $R : \mathfrak{g} \to \mathfrak{g}$ be a linear map. Let us define a bracket for all $X, Y \in \mathfrak{g}$ by

$$[X, Y]_R = [RX, Y] + [X, RY] . \tag{3.15}$$

It is clearly antisymmetric, but it will in general fail to satisfy the Jacobi identity. Indeed, if one works out the Jacobi identity for the new bracket one finds that this is equivalent to the following condition

$$[K(X, Y), Z] + [K(Y, Z), X] + [K(Z, X), Y] = 0 . \tag{3.16}$$

where $K : \bigwedge^2 \mathfrak{g} \to \mathfrak{g}$ is given by

$$K(X, Y) = R[RX, Y] + R[X, RY] - [RX, RY] . \tag{3.17}$$

A solution to (3.16) is provided by $K(X, Y) = [X, Y]$ in which case, (3.16) is satisfied by virtue of the Jacobi identity of the original Lie bracket on $\mathfrak{g}$. The resulting equation

$$R[RX, Y] + R[X, RY] - [RX, RY] = [X, Y] \tag{3.18}$$

is known as the *modified classical Yang–Baxter equation* (mCYBE).

In our case $R = \mathbb{P}_+ - \mathbb{P}_-$ and the new bracket $[-,-]_R$ is given simply by the Sklyanin bracket:

$$[X, Y]_R = 2[X_+, Y_+] - 2[X_-, Y_-] , \tag{3.19}$$

which can be checked to satisfy the Jacobi identity simply by using the properties of a Manin triple. We can also see this from (3.18) by noticing that in terms of $J = iR$, the mCYBE is equivalent to the vanishing of the Nijenhuis tensor (2.18) associated to $J$—that is, to the integrability of the complex structure. What we find curious is the fact that these two seemingly unrelated notions of integrability agree in this example. One may speculate on the possible relation between the quantization of the classical $r$-matrix and deformations of the complex structure; or more generally whether the quantum Yang–Baxter equation has a complex geometrical analogue. We hope to return to these intriguing issues elsewhere.

§4   Three-dimensional solvable complex Lie algebras

It follows from the correspondence between Manin triples and $N$=2 structures that in order to classify the latter ones, it is sufficient to classify the former ones. In this section we start the classification of the solvable Lie algebras admitting an $N$=2 structure. Already for the existence of the Sugawara $N$=1 construction, the Lie algebra must be self-dual. Self-dual Lie algebras are not classified, but there exists a structure theorem [**21**] that tells us how to obtain them in terms of the operations of direct sum and double extension (see below) from the simple Lie algebras and the one-dimensional Lie algebra. A closer look at this construction will tell us that the classification of $N$=2 structures with $c$=9 from solvable Lie algebras, is equivalent to the classification of six-dimensional solvable complex Manin triples. This will require in particular a knowledge of all the complex three-dimensional solvable Lie algebras.

Self-dual Lie algebras and double extensions

A double extension can be thought of as a machine to construct self-dual Lie algebras out of a pair $(\mathfrak{g}, \mathfrak{h})$, where $\mathfrak{g}$ is itself a self-dual Lie algebra, and $\mathfrak{h}$ is a Lie algebra acting on $\mathfrak{g}$ via derivations which also preserve the metric. We call such derivations antisymmetric. The notion of a double extension was introduced in [**21**] and it was investigated further in [**13**] in the context of Sugawara constructions, to where the reader is sent for a more explicit description.

Let $(\mathfrak{g}, \langle - , - \rangle_\mathfrak{g})$ be a self-dual Lie algebra and let $\mathfrak{h}$ be a Lie algebra acting on $\mathfrak{g}$ by antisymmetric derivations. Then the vector space $\mathfrak{g} \oplus \mathfrak{h} \oplus \mathfrak{h}^*$ can be endowed with a Lie bracket and with an invariant metric turning it into a self-dual Lie algebra. Let $\rho : \mathfrak{h} \to \text{Der} \, \mathfrak{g}$ denote the representation of $\mathfrak{h}$ as (antisymmetric) derivations of $\mathfrak{g}$. Because the action of $\mathfrak{h}$ preserves the metric on $\mathfrak{g}$, for all $h \in \mathfrak{h}$ and $x, y \in \mathfrak{g}$,

$$\langle \rho(h) \, x \, , \, y \rangle_\mathfrak{g} = - \langle x \, , \, \rho(h) \, y \rangle_\mathfrak{g} . \tag{4.1}$$

This means that we have a map $\beta : \bigwedge^2 \mathfrak{g} \to \mathfrak{h}^*$ defined by

$$\langle \beta(x \, , \, y), h \rangle = \langle \rho(h) \, x \, , \, y \rangle_\mathfrak{g} , \tag{4.2}$$

where the $\langle - , - \rangle$ on the left-hand side is the dual pairing between $\mathfrak{h}$ and $\mathfrak{h}^*$. Furthermore because $\rho(h)$ is a derivation, it follows that $\beta$ is a 2-cocycle, whence the bracket

$$[x,y] = [x,y]_{\mathfrak{g}} + \beta(x,y) \tag{4.3}$$

defines a central extension of $\mathfrak{g}$ by $\mathfrak{h}^*$. This makes $\mathfrak{g} \oplus \mathfrak{h}^*$ into a Lie algebra. Moreover $\mathfrak{h}$ acts on this Lie algebra in a natural way: on $\mathfrak{g}$ it acts via $\rho$ and on $\mathfrak{h}^*$ it acts via the coadjoint representation $\mathrm{ad}^*$, and $\beta$ is invariant under the combined action. Therefore we can consider the semidirect product of $\mathfrak{h}$ with this $\mathfrak{h}^*$-extension of $\mathfrak{g}$. In other words, if $h \in \mathfrak{h}$, $x \in \mathfrak{g}$ and $\alpha \in \mathfrak{h}^*$, we define

$$[h,x] = \rho(h)\, x \qquad \text{and} \qquad [h,\alpha] = \mathrm{ad}^*(h)\, \alpha\;, \tag{4.4}$$

whereas we leave the Lie bracket on $\mathfrak{h}$ unperturbed. We call the resulting algebra the double extension of $\mathfrak{g}$ by $\mathfrak{h}$ and we write it $\mathfrak{D}_\rho(\mathfrak{g},\mathfrak{h})$ or sometimes $\mathfrak{D}(\mathfrak{g},\mathfrak{h})$ if we do not require to specify $\rho$. We emphasize that in the data defining a double extension it is not just $\mathfrak{g}$ that enters, but $(\mathfrak{g}, \langle - , - \rangle_{\mathfrak{g}})$ even if the notation does not reflect this. Actually, the construction is impervious to changes of scale in the metric $\langle - , - \rangle_{\mathfrak{g}}$ whence only its conformal class enters.

There is more, however, and $\mathfrak{D}(\mathfrak{g},\mathfrak{h})$ is actually self-dual. Indeed, we can extend the invariant metric $\langle - , - \rangle_{\mathfrak{g}}$ on $\mathfrak{g}$ to all of $\mathfrak{D}(\mathfrak{g},\mathfrak{h})$ as follows. First we declare $\mathfrak{g}$ to be orthogonal to $\mathfrak{h} \oplus \mathfrak{h}^*$ and then we declare the metric on this latter subspace to be the following. If $h \oplus \alpha, h' \oplus \alpha' \in \mathfrak{h} \oplus \mathfrak{h}^*$, we put

$$\langle h \oplus \alpha\,,\, h' \oplus \alpha' \rangle \equiv \langle h\,,\, h' \rangle_{\mathfrak{h}} + \langle \alpha\,,\, h' \rangle + \langle \alpha'\,,\, h \rangle\;, \tag{4.5}$$

where $\langle -,- \rangle_{\mathfrak{h}}$ is any (possibly degenerate) invariant symmetric bilinear form on $\mathfrak{h}$.

A self-dual Lie algebra is called indecomposable if it cannot be written as the direct sum of orthogonal ideals. The main theorem of [**21**] says that an indecomposable self-dual Lie algebra is either simple, one-dimensional, or the double extension of a self-dual Lie algebra (possibly decomposable) by a simple or one-dimensional Lie algebra. Since a solvable Lie algebra cannot have a simple subalgebra, it follows that an indecomposable self-dual solvable Lie algebra is either one-dimensional or the double extension of a smaller solvable self-dual Lie algebra by the one-dimensional Lie algebra. In other words, the class of solvable self-dual Lie algebras is generated by the one-dimensional Lie algebra under the operations of direct sum and double extension by the one-dimensional Lie algebra.

Let us now take a closer look at double extensions $\mathfrak{D}_\rho(\mathfrak{g},\mathfrak{h})$ where $\mathfrak{g}$ is some solvable self-dual Lie algebra and $\mathfrak{h}$ is the one-dimensional Lie algebra. If we fix a basis $\langle x_i \rangle$ for $\mathfrak{g}$, the Lie brackets are specified by the structure constants $[x_i, x_j] = f_{ij}{}^k x_k$, and the invariant metric by the inner products $g_{ij} = \langle x_i\,,\, x_j \rangle$. Let $h$ denote the generator of $\mathfrak{h}$ and $h^*$ be the canonical dual generator for $\mathfrak{h}^*$. The action of $\mathfrak{h}$ on $\mathfrak{g}$ is determined by a matrix whose entries $\rho_i{}^j$ are defined by $\rho(h)\, x_i = \rho_i{}^j x_j$. This matrix is required to be a derivation: $f_{ij}{}^k \rho_k{}^\ell = f_{kj}{}^\ell \rho_i{}^k + f_{ijk}{}^\ell \rho_j{}^k$ and antisymmetric, in the sense that $\rho_{ij} \equiv \rho_i{}^k g_{kj} = -\rho_{ji}$. One can easily recover the Lie brackets and the metric of the double-extension in the basis $\langle x_i, h, h^* \rangle$ as follows. The nonzero brackets are given by

$$[x_i, x_j] = f_{ij}{}^k x_k + \rho_{ij} h^* \qquad \text{and} \qquad [h, x_i] = \rho_i{}^j x_j\;, \tag{4.6}$$

and the nonzero inner products are given by

$$\langle h\,,\, h^* \rangle = 1 \qquad \text{and} \qquad \langle x_i\,,\, x_j \rangle = g_{ij}\;, \tag{4.7}$$

whereas $\langle h\,,\, h \rangle$ is arbitrary.

We can now use this information to compute the Virasoro central charge (2.9) of the solvable double extension $\mathfrak{D}_\rho(\mathfrak{g},\mathfrak{h})$ defined above. A brief calculation shows that the central charge $c(\mathfrak{D})$ associated to the double extension is given by

$$c(\mathfrak{D}) = c(\mathfrak{g}) + 3\;. \tag{4.8}$$

Hence double extending by the one-dimensional Lie algebra only increases the Virasoro central charge by 3. (A similar result applies to any double extension [**13**].) But now $\mathfrak{g}$ is a self-dual solvable Lie algebra which, by the structure theorem, is a direct sum of double-extensions or of one-dimensional algebras. Applying the above argument again to each of the double extensions, we prove by induction on the dimension of the Lie algebra that the cental charge coming from a solvable Lie algebra $\mathfrak{s}$ is given by $c(\mathfrak{s}) = \frac{3}{2} \dim \mathfrak{s}$.

Therefore we see that the classification problem for the $N{=}2$ constructions with $c{=}9$ coming from complex solvable Lie algebras, is identical to the classification problem for six-dimensional complex solvable Lie algebras with an invariant metric and a compatible integrable complex structure. And from the results of the last section, this is equivalent to classifying the complex Manin triples $(\mathfrak{g}, \mathfrak{g}_+, \mathfrak{g}_-)$. If $\mathfrak{g}$ is solvable, so are $\mathfrak{g}_\pm$. The converse need not hold: one can have $\mathfrak{g}_\pm$ solvable but $\mathfrak{g}_{\mathbb{C}}$ not solvable (indeed, even semisimple). Hence to classify such Manin triples it is sufficient to classify the three-dimensional complex solvable (and cosolvable) Lie bialgebras, but we may then have to discard some.

Three-dimensional solvable Lie algebras

We now start this procedure by first classifying the complex solvable three-dimensional Lie algebras. Some of the Lie algebras in this section will be later complexified and will remain complexified for the rest of the paper. To relieve some of the notational burden that would ensue by having to write a $\mathfrak{g}_\mathbb{C}$ for the complexification of $\mathfrak{g}$, we prefer in this section when naming the real solvable Lie algebras, to adorn them with a $^\mathbb{R}$ and reserve the unadorned symbol for the complex Lie algebras.

The three-dimensional real Lie algebras were classified by Bianchi. Besides the abelian Lie algebra $\mathfrak{a}_3^\mathbb{R}$, there are four other three-dimensional real solvable Lie algebras (two of which have a parameter). They are all summarized in Table 1, which is partially borrowed from [27]. Notice that $\mathfrak{s}_{3,4}^\mathbb{R}(0) \cong \mathfrak{e}_2^\mathbb{R}$, the euclidean algebra; $\mathfrak{s}_{3,3}^\mathbb{R}(-1) \cong \mathfrak{e}_{1,1}^\mathbb{R}$, the pseudoeuclidean algebra in signature $(+,-)$; $\mathfrak{s}_{3,3}^\mathbb{R}(1) \cong \mathfrak{a}_1^\mathbb{R} \ltimes \mathfrak{t}_2^\mathbb{R}$, where $\mathfrak{t}_2^\mathbb{R}$ is the two-dimensional translation algebra and $\mathfrak{a}_1^\mathbb{R}$ acts as dilatations; and that $\mathfrak{s}_{3,3}^\mathbb{R}(0) \cong \mathfrak{a}_1^\mathbb{R} \times \mathfrak{s}_2^\mathbb{R}$, where $\mathfrak{s}_2^\mathbb{R}$ is the non-abelian two-dimensional Lie algebra. Also there exists a contraction in which $\lim_{\mu\to\infty} \mathfrak{s}_{3,4}^\mathbb{R}(\mu) \cong \mathfrak{s}_{3,3}^\mathbb{R}(1)$.

| $\mathfrak{g}$ | Nonzero Brackets | |
|---|---|---|
| $\mathfrak{a}_3^\mathbb{R}$ | | |
| $\mathfrak{s}_{3,1}^\mathbb{R}$ | $[e_2,e_3]=e_1$ | |
| $\mathfrak{s}_{3,2}^\mathbb{R}$ | $[e_1,e_3]=e_1$, | $[e_2,e_3]=e_1+e_2$ |
| $\mathfrak{s}_{3,3}^\mathbb{R}(\mu)$ | $[e_1,e_3]=e_1$, | $[e_2,e_3]=\mu e_2$ ($|\mu|\leq 1$) |
| $\mathfrak{s}_{3,4}^\mathbb{R}(\mu)$ | $[e_1,e_3]=\mu e_1-e_2$, | $[e_2,e_3]=e_1+\mu e_2$ ($\mu\geq 0$) |

**Table 1** *Real three-dimensional solvable Lie algebras*

Complex three-dimensional solvable Lie algebras

If we now think of these Lie algebras as complex Lie algebras, we find that some of them are isomorphic. In fact, let us elaborate a little bit on this. Every solvable three-dimensional Lie algebra $\mathfrak{g}$ has a two-dimensional abelian subalgebra $\mathfrak{h}$. Let us choose basis $\langle e_i \rangle$ for $\mathfrak{g}$ as in Table 1, such that $\mathfrak{h}$ is spanned by $e_1$ and $e_2$. Notice moreover that we can always choose the basis in such a way that the adjoint action of $e_3$ stabilizes $\mathfrak{h}$; that is, so that $e_3$ never appears in the right-hand side of $[e_i, e_3]$. Relative to this basis, $\mathfrak{g}$ is determined by the $2\times 2$ matrix $A$ defining the adjoint action of $e_3$ on $\mathfrak{h}$. What changes of basis are allowed while still keeping this structure? Let us first assume that we only allow changes of basis which preserve $\mathfrak{h}$. It is then clear that the most general such change of basis is $e_i' = T_{ij} e_j$ where the matrix $T$ has the form

$$\begin{pmatrix} S & 0 \\ * & s \end{pmatrix}$$

where $S$ is the matrix of an invertible transformation $\mathfrak{h} \to \mathfrak{h}$ and $s \neq 0$. It is clear that the matrix $A$ characterizing the algebra changes to $A \rightsquigarrow A' = sSAS^{-1}$. Thus we are looking at the orbit of $A$ under the action of conjugation by $GL(2)$ and multiplication by nonzero constants; or in other words, taking into account the freedom to rescale, $A$ defines a point in the projective space $\mathbb{P}^3$, and we are then asking for the $PSL(2)$ orbits in $\mathbb{P}^3$. The solution of this problem depends on the base field. Over the complex numbers we can always choose basis so that $A$ is in canonical Jordan form; that is, so that it agrees with one of the following matrices:

$$\begin{pmatrix} 0 & 0 \\ 0 & 0 \end{pmatrix}, \quad \begin{pmatrix} \alpha & 0 \\ 1 & \alpha \end{pmatrix}, \quad \text{or} \quad \begin{pmatrix} \alpha & 0 \\ 0 & \beta \end{pmatrix}. \tag{4.9}$$

If we now use the freedom to rescale, we can always choose $A$ to be one of the following:

$$\begin{pmatrix} 0 & 0 \\ 0 & 0 \end{pmatrix}, \quad \begin{pmatrix} 0 & 0 \\ 1 & 0 \end{pmatrix}, \quad \begin{pmatrix} 1 & 0 \\ 1 & 1 \end{pmatrix}, \quad \text{and} \quad \begin{pmatrix} 1 & 0 \\ 0 & \mu \end{pmatrix}, \tag{4.10}$$

where $\mu$ is a complex number satisfying $|\mu| \leq 1$, with the proviso that if $|\mu| = 1$ then we must further identify $\mu$ with $1/\mu$. We call the above forms of the matrix: *canonical projective Jordan forms*, since they take into account the rescaling of the matrix.

We are still not done, however. We must now study the possibility that two Lie algebras specified by different matrices are isomorphic via a transformation that does not preserve the abelian subalgebra $\mathfrak{h}$. Suppose that $\langle e_1, e_2 \rangle$ span $\mathfrak{h}$. We ask whether there exist commuting linearly independent vectors $e_1'$ and $e_2'$ of the form $e_a' = S_{ai} e_i$ where $a$ takes the values 1 and 2. The sufficient and necessary condition on $S$ is

$$(S_{13}S_{2a} - S_{1a}S_{23})A_{ab} = 0 . \tag{4.11}$$

We notice that $A$ cannot be singular, since the Lie algebras associated to the singular matrices in (4.10) cannot be isomorphic under complexification. Indeed, these algebras correspond to the the first two matrices in (4.10) and the last one with $\mu = 0$; and these Lie algebras have different structures: the Lie algebra associated to the first matrix is abelian; the one associated to the second matrix is nilpotent but nonabelian; whereas the algebra associated to the last matrix with $\mu = 0$ is solvable but not nilpotent. Thus we take $A$ nonsingular. In this case, (4.11) holds if and only if

$$S_{13}S_{2a} = S_{23}S_{1a} . \tag{4.12}$$

Notice that if $S_{13} = S_{23} = 0$ then we don't change $\mathfrak{h}$ and hence neither can we change the canonical projective Jordan form of the matrix $A$. Without loss of generality (that is, permuting the basis in $\mathfrak{h}$ if necessary) we can assume that $S_{13} \neq 0$. We can in fact, take it to be 1 and take $S_{23} = \nu$. Then the condition (4.12) simply says that $S_{2a} = \nu S_{1a}$. But this violates linear independence. Indeed, if this is the case, the $2 \times 3$ matrix for $S$ is given by

$$S = \begin{pmatrix} S_{11} & S_{12} & S_{13} \\ S_{21} & S_{22} & S_{23} \end{pmatrix} = \begin{pmatrix} S_{11} & S_{12} & 1 \\ \nu S_{11} & \nu S_{12} & \nu \end{pmatrix} , \tag{4.13}$$

which clearly has rank only 1. We therefore conclude that $S_{13} = 0$. Similarly, the same argument says that $S_{23} = 0$, whence $\mathfrak{h}$ is preserved. Therefore there is no further freedom in changing basis and isomorphism classes of complex three-dimensional solvable Lie algebras are in one-to-one correspondence with $2 \times 2$ matrices in canonical projective Jordan form.

This classification is summarized in Table 2. Comparing with the classification of real three-dimensional solvable Lie algebras, notice that $\mathfrak{s}_{3,4}^{\mathbb{R}}(\mu)$ disappears from the classification. In fact, we'll see below that it is a real form of the complexification of $\mathfrak{s}_{3,3}^{\mathbb{R}}(\mu')$ with $\mu' = \frac{\mu-i}{\mu+i}$, the Cayley transform of $\mu$.

| $\mathfrak{g}$ | Nonzero Brackets | |
|---|---|---|
| $\mathfrak{a}_3$ | | |
| $\mathfrak{s}_{3,1}$ | $[e_2, e_3] = e_1$ | |
| $\mathfrak{s}_{3,2}$ | $[e_1, e_3] = e_1$ , | $[e_2, e_3] = e_1 + e_2$ |
| $\mathfrak{s}_{3,3}(\mu)$ | $[e_1, e_3] = e_1$ , | $[e_2, e_3] = \mu e_2$ $\quad (|\mu| \leq 1)$ |

**Table 2** *Complex three-dimensional solvable Lie algebras*

If $|\mu| = 1$, we have in addition the isomorphism $\mathfrak{s}_{3,3}(\mu) \cong \mathfrak{s}_{3,3}(1/\mu)$.

In fact, unlike semisimple Lie algebras, a complex solvable Lie algebra need not be the complexification of a real Lie algebra as justified in the previous section using deformation theory arguments. We can see this explicitly from the above classification. The algebra $\mathfrak{s}_{3,3}(\mu)$ is the complexification of a real algebra only when the parameter $\mu$ is either real in which case it is the complexification of $\mathfrak{s}_{3,3}^{\mathbb{R}}(\mu)$; or when $|\mu| = 1$, in which case it is a complexification of $\mathfrak{s}_{3,4}^{\mathbb{R}}(\mu')$ with $\mu'$ essentially the Cayley transform of $\mu$. This can be seen as follows. The question is which projective Jordan canonical forms are congruent (in $GL(2, \mathbb{C})$) to a complex multiple of a real matrix. It is clear that the first three matrices in (4.10) are already real and so is the fourth when $\mu \in \mathbb{R}$. How about then when $\mu$ is complex? We can use the freedom to rescale by a complex number to turn this matrix into a general diagonal matrix with eigenvalues $\lambda$ and $\nu = \lambda\mu$. The question is then whether this is congruent in $GL(2, \mathbb{C})$ to a real matrix. Now congruence preserves the characteristic polynomial, which for a real matrix is real. This means that $(t - \lambda)(t - \nu)$ is a real polynomial. Thus either the conjugation fixes the roots—hence $\lambda$ and $\nu$ are real—or else permutes them—whence they are complex conjugate. In the former case, $\mu$ is real; and in the latter we find $\mu = \overline{\lambda}/\lambda$, so that $|\mu| = 1$. Conversely, let $|\mu| = 1$ but not real; then the last matrix in (4.10) is congruent to a multiple of

$$\begin{pmatrix} \rho & -1 \\ 1 & \rho \end{pmatrix} \tag{4.14}$$

with $\rho = i\frac{1+\mu}{1-\mu}$, the inverse Cayley transform of $\mu$.

Automorphisms

In the sequel we will need the explicit form of the automorphism groups of the Lie algebras in Table 2. These are the Lie subgroups of $GL(3, \mathbb{C})$ which preserve the Lie brackets. We expect relatively large automorphism groups, since contrary to the case of semisimple Lie algebras, solvable Lie algebras generically have outer automorphisms. The automorphism groups are listed in Table 3. For the abelian algebra, any invertible transformation is an automorphism; but we treat the other cases separately.

- Case: $\mathfrak{s}_{3,1}$.

With $\{e_i\}$ obeying the Lie brackets of $\mathfrak{s}_{3,1}$, we define $e_i' = \sum_j e_j A_{ji}$ and we demand that $[e_2', e_3'] = e_1'$ and all other brackets be zero. This leads to the following equations on the coefficients $A_{ij}$:

$$\begin{aligned} A_{21}A_{32} - A_{31}A_{22} &= 0 \\ A_{21}A_{33} - A_{31}A_{23} &= 0 \\ A_{22}A_{33} - A_{32}A_{23} &= A_{11} \\ A_{21} = A_{31} &= 0 . \end{aligned} \tag{4.15}$$

| $\mathfrak{g}$ | Automorphism Group |
|---|---|
| $\mathfrak{a}_3$ | $GL(3,\mathbb{C})$ |
| $\mathfrak{s}_{3,1}$ | $\begin{pmatrix} \det A & v^t \\ 0 & A \end{pmatrix}, \quad A \in GL(2,\mathbb{C}), \quad v \in \mathbb{C}^2$ |
| $\mathfrak{s}_{3,2}$ | $\begin{pmatrix} a & b & c \\ 0 & a & d \\ 0 & 0 & 1 \end{pmatrix}, \quad a \in \mathbb{C}^\times, \quad b,c,d \in \mathbb{C}$ |
| $\mathfrak{s}_{3,3}(1)$ | $\begin{pmatrix} A & v \\ 0 & 1 \end{pmatrix}, \quad A \in GL(2,\mathbb{C}), \quad v \in \mathbb{C}^2$ |
| $\mathfrak{s}_{3,3}(-1)$ | $\begin{pmatrix} a & 0 & v_1 \\ 0 & b & v_2 \\ 0 & 0 & 1 \end{pmatrix}, \begin{pmatrix} 0 & a & v_1 \\ b & 0 & v_2 \\ 0 & 0 & -1 \end{pmatrix}, a,b \in \mathbb{C}^\times, v_i \in \mathbb{C}$ |
| $\mathfrak{s}_{3,3}(\mu \neq \pm 1)$ | $\begin{pmatrix} A & v \\ 0 & 1 \end{pmatrix}, \quad A = \begin{pmatrix} a & 0 \\ 0 & b \end{pmatrix} \in GL(2,\mathbb{C}), \quad v \in \mathbb{C}^2$ |

**Table 3** *Automorphism groups of the Lie algebras in Table 2*

The most general solution to these equations is

$$(A_{ij}) = \begin{pmatrix} \det A & v^t \\ 0 & A \end{pmatrix}, \qquad (4.16)$$

where $A$ is in $GL(2,\mathbb{C})$ and $v \in \mathbb{C}^2$.

• Case: $\mathfrak{s}_{3,2}$.

We now take $\{e_i\}$ to obey the Lie brackets of $\mathfrak{s}_{3,2}$ and we define $e'_i = \sum_j e_j A_{ji}$. Imposing that they too satisfy the nonzero Lie brackets $[e'_1, e'_3] = e'_1$ and $[e'_2, e'_3] = e'_1 + e'_2$, we find the following equations:

$$\begin{aligned}
A_{11}A_{32} - A_{31}A_{12} &= 0 \\
A_{21}A_{32} - A_{31}A_{22} &= 0 \\
A_{11}A_{33} - A_{31}A_{13} + A_{21}A_{33} - A_{31}A_{23} &= A_{11} \\
A_{21}A_{33} - A_{31}A_{23} &= A_{21} \\
A_{31} &= 0 \\
A_{12}A_{33} - A_{32}A_{13} + A_{22}A_{33} - A_{32}A_{23} &= A_{11} + A_{12} \\
A_{22}A_{33} - A_{32}A_{23} &= A_{21} + A_{22} \\
A_{31} + A_{32} &= 0 \, .
\end{aligned} \qquad (4.17)$$

Clearly then $A_{31} = A_{32} = 0$. Nondegenracy of $A_{ij}$ implies that $A_{33} = 1$, whence $A_{21} = 0$ and $A_{11} = A_{22}$. In brief, the most general solution is then

$$(A_{ij}) = \begin{pmatrix} a & b & c \\ 0 & a & d \\ 0 & 0 & 1 \end{pmatrix}, \qquad (4.18)$$

where $a,b,c,d \in \mathbb{C}$ and $a \neq 0$.

• Case: $\mathfrak{s}_{3,3}(\mu)$.

The equations in this case are the following (after some massaging):

$$\begin{aligned}
A_{31} &= 0 \\
\mu A_{32} &= 0 \\
A_{11}A_{32} &= 0 \\
A_{11}(A_{33} - 1) &= 0 \\
A_{21}(\mu A_{33} - 1) &= 0 \\
A_{12}(A_{33} - \mu) &= A_{32}A_{13} \\
\mu A_{22}(A_{33} - 1) &= 0 \, .
\end{aligned} \qquad (4.19)$$

We must distinguish three cases: $\mu = 1$, $\mu = -1$, and $\mu \neq \pm 1$. The case $\mu = 0$ is actually generic, contrary to one's naive expectation.

◦ Subcase: $\mathfrak{s}_{3,3}(1)$.

Nondegeneracy of $A_{ij}$ forces $A_{33} = 1$ in this case and the general solution is

$$(A_{ij}) = \begin{pmatrix} A & v^t \\ 0 & 1 \end{pmatrix}, \qquad (4.20)$$

where $A$ is any nondegenerate $2 \times 2$ matrix, and $v \in \mathbb{C}^2$.

○ Subcase: $\mathfrak{s}_{3,3}(-1)$.

In this case, there are two possibilities: $A_{33} = 1$, in which case $A_{21} = A_{12} = 0$; or $A_{33} = -1$, which forces $A_{11} = A_{22} = 0$. In other words, the automorphism group has two connected components:

$$(A_{ij}) = \begin{pmatrix} a & 0 & v_1 \\ 0 & b & v_2 \\ 0 & 0 & 1 \end{pmatrix} \quad \text{or} \quad (A_{ij}) = \begin{pmatrix} 0 & a & v_1 \\ b & 0 & v_2 \\ 0 & 0 & -1 \end{pmatrix}, \tag{4.21}$$

where $a, b, v_i \in \mathbb{C}$ and $a, b \neq = 0$.

○ Subcase: $\mathfrak{s}_{3,3}(\mu \neq \pm 1)$.

The unique solution is $A_{33} = 1$, which implies $A_{12} = A_{21} = 0$. Hence we only get the connected component of the identity of the previous case:

$$(A_{ij}) = \begin{pmatrix} a & 0 & v_1 \\ 0 & b & v_2 \\ 0 & 0 & 1 \end{pmatrix}, \tag{4.22}$$

where $a, b \in \mathbb{C}^\times$ and $v_i \in \mathbb{C}$.

## §5  Three-dimensional (co)solvable complex Lie bialgebras

Having classified the three-dimensional solvable complex Lie algebras it is now a simple matter—albeit somewhat tedious—to check which Lie algebras in Table 2 are bialgebras. We can do this in two different ways. One way would be simply to fix a basis and the brackets as in the table and solve the cocycle condition (3.10) for the cobracket. We then fix the remaining freedom by the Jacobi identity for the cobracket. Of course, the brackets obtained for $\mathfrak{g}^*$ will not necessarily agree with the ones in Table 2. This is because the basis we have chosen for $\mathfrak{g}^*$ is one which is canonically dual to the chosen basis for $\mathfrak{g}$ and not necessarily the one chosen in Table 2. Finding the right basis is a tedious task and we prefer to use this method only to test for the existence of a pairing. To actually determine the pairings, we prefer an alternative method which is to list the solvable three-dimensional Lie bialgebras as *matched* pairs of Lie algebras. Two Lie algebras $\mathfrak{g}_1$, $\mathfrak{g}_2$ are said to be matched if their vector space direct sum admits the structure of a self-dual Lie algebra in such a way that $\mathfrak{g}_1$ and $\mathfrak{g}_2$ are isotropic subalgebras. The resulting self-dual Lie algebra is usually denoted $\mathfrak{g}_1 \bowtie \mathfrak{g}_2$, although it should be kept in mind that even though the notation does not reflect it, two Lie algebras can be paired in more than one way.

Our approach will thus be the following: we will choose $\mathfrak{g}_1$ and $\mathfrak{g}_2$ to be Lie algebras in Table 2 with a fixed set of generators $\{e_i^{(1)}\}$ and $\{e_i^{(2)}\}$ respectively, and with an invariant metric given by $\langle e_i^{(1)}, e_j^{(2)} \rangle = g_{ij}$ and $\langle e_i^{(1)}, e_j^{(1)} \rangle = \langle e_i^{(2)}, e_j^{(2)} \rangle = 0$; where $g_{ij}$ is not necessarily symmetric, but rather $\langle -, - \rangle$ is. Given this data, we can recover the Lie algebraic structure of $\mathfrak{g}_1 \bowtie \mathfrak{g}_2$ as follows: $\mathfrak{g}_1$ and $\mathfrak{g}_2$ are embedded as subalgebras; and the mixed brackets are given by

$$[e_i^{(1)}, e_j^{(2)}] = f_{j\ell}^{(2)m} g_{im} g^{\ell k} e_k^{(1)} - f_{i\ell}^{(1)m} g_{mj} g^{k\ell} e_k^{(2)}, \tag{5.1}$$

where $f_{ij}^{(1)k}$ and $f_{ij}^{(2)k}$ are structure constants of $\mathfrak{g}_1$ and $\mathfrak{g}_2$, respectively. Thus the Lie brackets depend explicitly on (the conformal class of) the pairing $g_{ij}$ and it might happen that different pairings will yield nonisomorphic Lie algebras.

The strategy will then be the following:

(1) We first analyze in turn each pair $(\mathfrak{g}_1, \mathfrak{g}_2)$ of solvable three-dimensional complex Lie algebras, and we determine if and how they can be paired; that is, we determine all the possible nondegenerate pairings $g_{ij}$ such that the mixed brackets (5.1) satisfy the Jacobi identity.

(2) For each matchable pair of algebras, we determine those pairings which give rise to nonisomorphic matched pairs.

(3) Finally we determine which of the isomorphism classes of matched pairs are actually solvable.

In this section we describe the first of these three steps. We summarize our results in Table 4 below, which contains the matched pairs of three-dimensional complex solvable Lie algebras. Notice that if $\mathfrak{g}_1 \bowtie \mathfrak{g}_2$ is a matched pair, so is $\mathfrak{g}_2 \bowtie \mathfrak{g}_1$; so that we do not have to consider ordered pairs. Whenever a matched pair $\mathfrak{g}_1 \bowtie \mathfrak{g}_2$ appears, we read off the generators and brackets of $\mathfrak{g}_i$ from Table 2, and the mixed brackets are given by (5.1) and the expression for $g_{ij}$ in Table 4. The conditions in Table 4 on the possible $g_{ij}$ are rather weak. However not all of these different choices of $g_{ij}$ will lead to nonisomorphic matched pairs. We still have to take into account the fact that a basis change relates different $g_{ij}$. We will tackle this in the next section.

The notation in Table 4 is such that $\mathfrak{g}$ stands for any of the Lie algebras in Table 2. The details behind this classification are discussed below. Here we simply mention that it is not sufficient that $\mathfrak{g}_1$ and $\mathfrak{g}_2$ be solvable for $\mathfrak{g}_1 \bowtie \mathfrak{g}_2$ to be solvable. Requiring that $\mathfrak{g}_1 \bowtie \mathfrak{g}_2$ be solvable, will restrict the list further. We will also see this in the next section.

| Matched Pair | Admissible Pairing ($g_{ij}$) |
|---|---|
| $\mathfrak{a}_3 \bowtie \mathfrak{g}$ | $\det g \neq 0$ |
| $\mathfrak{s}_{3,1} \bowtie \mathfrak{s}_{3,1}$ | $\det g \neq 0$ and $g_{11} = 0$ |
| $\mathfrak{s}_{3,1} \bowtie \mathfrak{s}_{3,2}$ | $\det g \neq 0$ and $g_{11} = 0$ |
| $\mathfrak{s}_{3,1} \bowtie \mathfrak{s}_{3,3}(1)$ | $\det g \neq 0$ |
| $\mathfrak{s}_{3,1} \bowtie \mathfrak{s}_{3,3}(\mu \neq 1)$ | $\det g \neq 0$ and $g_{11} = g_{12} = 0$ |
| $\mathfrak{s}_{3,2} \bowtie \mathfrak{s}_{3,3}(-1)$ | $\det g \neq 0$ and $g_{11} = g_{12} = 0$ |
| $\mathfrak{s}_{3,3}(1) \bowtie \mathfrak{s}_{3,3}(-1)$ | $\det g \neq 0$ |
| $\mathfrak{s}_{3,3}(\mu \neq \pm 1) \bowtie \mathfrak{s}_{3,3}(-\mu)$ | $\det g \neq 0$ and $g_{12} = g_{21} = 0$ |

**Table 4** *Matched pairs of complex three-dimensional solvable Lie algebras*

<u>The details behind Table 4</u>

In this subsection we write down the details of the classification of complex solvable six-dimensional Manin triples summarized in Table 4. First of all notice that for any Lie algebra $\mathfrak{g}$ in Table 2, $\mathfrak{g} \bowtie \mathfrak{a}_3$ is a matched pair. We now take one by one the Lie algebras in Table 2 and check what algebras can be matched with them. Two ways present themselves to tackle this problem. In the first approach, we solve the cocycle condition (3.10) and the Jacobi identity for the cobracket in a basis dual to the basis chosen for the first algebra. In order to recognize the algebra as one of the ones in Table 2, we have to get it to a standard form; that is, writing $\mathfrak{g} = \mathfrak{h} \oplus \langle x \rangle$ where $\mathfrak{h}$ is an abelian two-dimensional subalgebra closed under the adjoint action of $x$, and such that the $2 \times 2$ matrix defining this action is in one of the canonical projective Jordan forms in (4.10). Alternatively, we can try to match two algebras from Table 2 by solving the cocycle condition (3.10) for the metric $g_{ij}$—the determining condition being that the metric be nondegenerate. Needless to say, we have benefitted from the use of *Mathematica* to check many of these identities. The programs and some of the sessions are available upon request.

• Matching $\mathfrak{g}_1 = \mathfrak{s}_{3,1}$.

With respect to a dual basis $\langle e^i \rangle$, $\mathfrak{g}_2$ is given by

$$[e^1, e^2] = -c^{23}{}_3 e^1 + c^{12}{}_2 e^2 + c^{12}{}_3 e^3$$
$$[e^1, e^3] = c^{23}{}_2 e^1 + c^{13}{}_2 e^2 + c^{13}{}_3 e^3 \quad (5.2)$$
$$[e^2, e^3] = c^{23}{}_2 e^2 + c^{23}{}_3 e^3$$

subject to the following conditions coming from the Jacobi identity

$$(c^{12}{}_2 - c^{13}{}_3)c^{23}{}_3 = 2c^{12}{}_3 c^{23}{}_2$$
$$(c^{13}{}_3 - c^{12}{}_2)c^{23}{}_2 = 2c^{13}{}_2 c^{23}{}_3 . \quad (5.3)$$

Notice that the Jacobi identity is automatically satisfied if we choose $c^{23}{}_3 = c^{23}{}_2 = 0$, which moreover turns $\langle e^2, e^3 \rangle$ into an abelian subalgebra closed under $[e^1, -]$. The matrix defining this action is moreover not constrained at all, whence all algebras are matched. We now determine the allowed pairings $g_{ij}$.

○ $\mathfrak{s}_{3,1} \bowtie \mathfrak{s}_{3,1}$

Imposing the mixed Jacobi identities, we find only two constraints on the entries $g_{ij}$: $g_{11} g_{21} = 0$ and $g_{11} = 0$.

○ $\mathfrak{s}_{3,1} \bowtie \mathfrak{s}_{3,2}$

Just as before the two conditions are $g_{11} g_{21} = 0$ and $g_{11} = 0$.

○ $\mathfrak{s}_{3,1} \bowtie \mathfrak{s}_{3,3}(\mu)$

This divides itself into two cases: $\mu = 1$ and $\mu \neq 1$. For $\mu = 1$ it is easy to see that there are no conditions on $g_{ij}$, hence any pairing will do. For $\mu \neq 1$ we find the following conditions: $g_{11} g_{12} = g_{11} g_{22} = g_{11} g_{32} = g_{12} g_{21} = g_{12} g_{31} = 0$. If $g_{11} \neq 0$ then $g_{12} = g_{22} = g_{32} = 0$, whence $(g_{ij})$ would be degenerate. Hence $g_{11} = 0$. Similarly, if $g_{12} \neq 0$, then it would force $g_{21} = g_{31} = 0$ which, since $g_{11} = 0$, would force $(g_{ij})$ to be degenerate. Thus, $g_{12} = 0$.

• Matching $\mathfrak{g}_1 = \mathfrak{s}_{3,2}$.

With respect to a dual basis $\langle e^i \rangle$, $\mathfrak{g}_2$ is given by

$$[e^1, e^2] = -c^{13}{}_3 e^2 + c^{12}{}_3 e^3$$
$$[e^1, e^3] = c^{13}{}_2 e^2 + c^{13}{}_3 e^3 \quad (5.4)$$
$$[e^2, e^3] = 0 .$$

Already we have $\langle e^2, e^3 \rangle$ as the two-dimensional abelian subalgebra stabilized by $e^1$. The $2 \times 2$ matrix characterizing this algebra is clearly traceless. Since tracelessness in an invariant property under rescalings and changes of basis in $\mathfrak{h}$, only those traceless matrices in (4.10) can correspond to $\mathfrak{g}_2$. Thus $\mathfrak{g}_2$ can be $\mathfrak{a}_3$, $\mathfrak{s}_{3,1}$ or $\mathfrak{s}_{3,3}(-1)$. We already knew the former two cases, but the latter is a new possibility. Indeed, choosing $c^{12}{}_3 = c^{13}{}_2 = 0$ and $c^{13}{}_3 = 1$, we obtain $\mathfrak{s}_{3,3}(-1)$, after the permutation of basis: $e^1 \leadsto e_3$, $e^2 \leadsto e_2$ and $e^3 \leadsto e_1$. We now find out the allowed pairings.

∘ $\mathfrak{s}_{3,2} \bowtie \mathfrak{s}_{3,3}(-1)$

The independent equations coming from the Jacobi identity for the mixed bracket are the following: $g_{11}g_{21} = g_{11}g_{22} = g_{11}g_{23} = g_{12}g_{21} = g_{13}g_{21} = 0$. It is clear that taking into account the nondegeneracy of $(g_{ij})$, the unique solution to these equations is $g_{11} = g_{21} = 0$.

• Matching $\mathfrak{g}_1 = \mathfrak{s}_{3,3}(\mu)$.

Here we find it easier to start with the second approach. Since we have already discussed all algebras in Table 2 except for $\mathfrak{s}_{3,3}(\mu)$, it is clear that all matched pairs have been taken into account except for those of the form $\mathfrak{s}_{3,3}(\mu) \bowtie \mathfrak{s}_{3,3}(\nu)$. We therefore set out to investigate if there exists a nondegenerate invariant metric pairing up these two algebras. The cocycle condition can be used to obtain equations (listed below) for the entries $g_{ij}$ of the metric. These equations naturally depend on $\mu$ and $\nu$ and it is then a matter of simple algebra to decide for which values of $\mu$ and $\nu$, these equations can be satisfied by a nondegenerate metric.

The mixed Jacobi identity for $\mathfrak{s}_{3,3}(\mu) \bowtie \mathfrak{s}_{3,3}(\nu)$, yields the following fifteen equations

$$0 = (1 + \mu\nu)g_{12}g_{31} - (\mu + \nu)g_{11}g_{32} \qquad (a1)$$
$$0 = (1 + \mu\nu)g_{12}g_{21} - (\mu + \nu)g_{11}g_{22} \qquad (a2)$$
$$0 = (1 + \mu\nu)g_{21}g_{32} - (\mu + \nu)g_{22}g_{31} \qquad (a3)$$
$$0 = (1 - \mu)(1 - \nu)g_{11}g_{12}g_{21} \qquad (b1)$$
$$0 = (1 - \mu)(1 - \nu)g_{11}g_{12}g_{22} \qquad (b2)$$
$$0 = (1 - \mu)(1 - \nu)g_{11}g_{12}g_{23} \qquad (b3)$$
$$0 = (1 - \mu)(1 - \nu)g_{11}g_{21}g_{22} \qquad (b4)$$
$$0 = (1 - \mu)(1 - \nu)g_{12}g_{21}g_{22} \qquad (b5)$$
$$0 = (1 - \mu)(1 - \nu)g_{13}g_{21}g_{22} \qquad (b6)$$
$$0 = (\mu + \nu)g_{11}(g_{12}g_{31} - g_{11}g_{32}) \qquad (c1)$$
$$0 = (\mu + \nu)g_{22}(g_{21}g_{32} - g_{22}g_{31}) \qquad (c2)$$
$$0 = (1 + \mu\nu)g_{12}(g_{12}g_{31} - g_{11}g_{32}) \qquad (d1)$$
$$0 = (1 + \mu\nu)g_{21}(g_{22}g_{31} - g_{21}g_{32}) \qquad (d2)$$
$$0 = (\mu + \nu)g_{11}g_{13}g_{32} + (1 - \mu)(1 - \nu)g_{11}g_{12}g_{31} - (1 + \mu\nu)g_{12}g_{13}g_{31} \qquad (e1)$$
$$0 = (\mu + \nu)g_{22}g_{23}g_{31} + (1 - \mu)(1 - \nu)g_{21}g_{22}g_{33} - (1 + \mu\nu)g_{21}g_{23}g_{32} \ . \qquad (e2)$$

These equations clearly depend on the values of $\mu$ and $\nu$ through three polynomials $(1 + \mu\nu)$, $(1 - \mu)(1 - \nu)$, and $(\mu + \nu)$. We will therefore break up the problem in different cases depending on whether these polynomials vanish or not. Remember that both $\mu$ and $\nu$ are complex numbers inside the unit disk (or on the boundary) and that we can moreover choose $|\mu| \leq |\nu| \leq 1$, without loss of generality.

∘ Case: $\mu = \nu = 1$ (hence $\mu + \nu \neq 0$ and $1 + \mu\nu \neq 0$).

In this case, the first three equations (a1-3) imply the other twelve. These three equations imply the vanishing of three minors of $(g_{ij})$. If we now compute the determinant of $g_{ij}$ expanding along the third column, we find

$$\left\| \begin{matrix} g_{11} & g_{12} & g_{13} \\ g_{21} & g_{22} & g_{23} \\ g_{31} & g_{32} & g_{33} \end{matrix} \right\| = g_{13} \left\| \begin{matrix} g_{21} & g_{22} \\ g_{31} & g_{32} \end{matrix} \right\| - g_{23} \left\| \begin{matrix} g_{11} & g_{12} \\ g_{31} & g_{32} \end{matrix} \right\| + g_{33} \left\| \begin{matrix} g_{11} & g_{12} \\ g_{21} & g_{22} \end{matrix} \right\| , \qquad (5.5)$$

which vanishes due to the vanishing of each of the minor determinants. Therefore there is no matched pair with $\mu = \nu = 1$. Notice that this argument is also valid for $\mu = 1$ and $\nu \neq -1$.

∘ Case: $\mu + \nu = 0$, $\mu^2 = 1$.

In this case all the equations are automatically satisfied since when $\mu + \nu = 0$, they all come multiplied by $(1 - \mu^2)$. Hence there are no conditions on the pairing and any nondegenerate $(g_{ij})$ will do.

∘ Case: $\mu + \nu = 0$, $\mu^2 \neq 1$.

Notice that $(1 + \mu\nu) = (1 - \mu)(1 - \nu) \neq 0$. Notice then that equations (b1-3) imply that $g_{11}g_{12} = 0$ if $(g_{ij})$ is to avoid having a zero row. But this together with (a1-2) imply that $g_{12} = 0$ for $(g_{ij})$ not to have a zero column. Similarly, equations (b4-6) imply that $g_{21}g_{22} = 0$ lest $(g_{ij})$ develop a zero row, and this and (a3) together imply that $g_{21} = 0$, or $(g_{ij})$ has a zero column. Once $g_{12} = g_{21} = 0$ all the other equations are automatically satisfied.

∘ Case: $\mu + \nu \neq 0$, $1 + \mu\nu \neq 0$, $\mu \neq 1$, $\nu \neq 1$.

Notice now, that because $g_{ij}$ cannot have a zero row, equations (b1-3) say that $g_{11}g_{12} = 0$; and similarly equations (b4-6) say that $g_{21}g_{22} = 0$. Plugging these formulae into equations (c1-2), we find that also $g_{11}g_{32} = g_{22}g_{31} = 0$. Because $1 + \mu\nu \neq 0$, we see from (a1) and (a3) that therefore $g_{12}g_{31} = 0$ and $g_{21}g_{32} = 0$, whereas from (a2) we find that

$$g_{11}g_{22} = \lambda g_{12}g_{21} \qquad (5.6)$$

where $\lambda \equiv \frac{1+\mu\nu}{\mu+\nu} \neq 0$. Squaring this relation we find

$$(g_{11}g_{22})^2 = \lambda g_{11}g_{22}g_{12}g_{21} , \qquad (5.7)$$

which vanishes since $g_{21}g_{22} = 0$. Therefore $g_{11}g_{22} = 0$ and hence by (5.6), also

$g_{12}g_{21} = 0$. In summary the following monomials all vanish:

$$g_{11}g_{12} \quad g_{11}g_{22} \quad g_{11}g_{32} \quad g_{12}g_{21} \quad g_{12}g_{31} \quad g_{21}g_{22} \quad g_{21}g_{32} \quad g_{22}g_{31} \ . \tag{5.8}$$

Then plugging this into the expansion of the determinant of the metric given in (5.5), we find that all the minor determinants vanish and hence the matrix $(g_{ij})$ is degenerate. In other words, there is no matched pair.

∘ Case: $\mu + \nu \neq 0$, $\mu\nu = -1$, $\mu \neq 1$, $\nu \neq 1$.

This is the final case. Arguing as before we find that the cocycle equations are equivalent to the vanishing of the following monomials:

$$g_{11}g_{12} \quad g_{21}g_{22} \quad g_{11}g_{32} \quad g_{22}g_{31} \quad g_{11}g_{22} \ . \tag{5.9}$$

Since the metric cannot have a zero column, the unique admissible solution to these equations is $g_{11} = g_{22} = 0$. Notice, however, that this is the same (after a trivial change of variable) to the case $\mu + \nu = 0$, $\mu \neq 1$, $\nu \neq 1$; since as we saw in Table 2, $\mathfrak{s}_{3,3}(1/\mu) \cong \mathfrak{s}_{3,3}(\mu)$ for $|\mu| = 1$. In fact, the isomorphism involves transposing $e_1$ and $e_2$ and rescaling $e_3$. As a check of the above, we notice the conditions on $g$ found here agree with those of that case after the change of variables.

## §6  The moduli space of solvable $N$=2 structures

In the previous section we have classified all possible matched pairs $\mathfrak{g}_1 \bowtie \mathfrak{g}_2$, where $\mathfrak{g}_1$ and $\mathfrak{g}_2$ are complex three-dimensional solvable Lie algebras and the result is given by Table 4. The data determining each matched pair in Table 4 is a triple $(f_{ij}^{(1)k}, f_{ij}^{(2)k}, g_{ij})$, where $f_{ij}^{(1)k}$ (resp. $f_{ij}^{(2)k}$) are the structure constants of $\mathfrak{g}_1$ (resp. $\mathfrak{g}_2$) relative to a fixed basis $\{e_i^{(1)}\}$ (resp. $\{e_i^{(2)}\}$); and $g_{ij}$ are the entries of a matrix in $GL(3, \mathbb{C})$ satisfying in addition the conditions in Table 4.

In fact, only the image $[g_{ij}]$ of $g_{ij}$ in $PGL(3, \mathbb{C})$ enters in the determination of the Lie bracket of the matched pair. This follows from (5.1), which shows that the Lie brackets of the matched pair only depend on the conformal class of $g_{ij}$. Of course, rescaling $g_{ij}$ will rescale the metric of the Lie algebra $\mathfrak{g}_1 \bowtie \mathfrak{g}_2$ which will result in a rescaling of the $N$=2 generators. Nevertheless many important properties of the $N$=2 theory (for instance, the chiral ring) are invariant under such a rescaling; and for the purposes of this section, we will treat $N$=2 structures which differ only by a rescaling of the invariant metric as equivalent. In fact, we will see that in many cases this does not lead to any loss of generality, since $N$=2 structures which are equivalent in this weaker sense are strictly isomorphic. This will be the case, in fact, for the solvable cases;

hence the classification we will arrive at will be a classification of strictly nonisomorphic $N$=2 structures. In any case, even when we take this rescaling into consideration, different data do not generally lead to nonisomorphic matched pairs and the purpose of this section is to narrow down the list in Table 4 to only those matched pairs which give rise to nonisomorphic $N$=2 structures.

### Nonisomorphic matched pairs

We have seen in Section 3 that $N$=2 structures are in bijective correspondence with Manin triples $(\mathfrak{g}_1 \bowtie \mathfrak{g}_2, \mathfrak{g}_1, \mathfrak{g}_2)$. And two Manin triples $(\mathfrak{g}_1 \bowtie \mathfrak{g}_2, \mathfrak{g}_1, \mathfrak{g}_2)$ and $(\mathfrak{g}_1' \bowtie \mathfrak{g}_2', \mathfrak{g}_1', \mathfrak{g}_2')$ are isomorphic when there is an isomorphism $\mathfrak{g}_1 \bowtie \mathfrak{g}_2 \xrightarrow{\sim} \mathfrak{g}_1' \bowtie \mathfrak{g}_2'$ which restricts to Lie algebra isomorphisms $\mathfrak{g}_1 \xrightarrow{\sim} \mathfrak{g}_1'$ and $\mathfrak{g}_2 \xrightarrow{\sim} \mathfrak{g}_2'$. Very concretely, in terms of the data given in Table 4, we are looking for two Manin triples with data $(f_{ij}^{(1)k}, f_{ij}^{(2)k}, [g_{ij}])$ and $(f_{ij}^{(1)k}, f_{ij}^{(2)k}, [g_{ij}'])$, which are related by a change of basis of the ambient vector space spanned by $\{e_i^{(1)}, e_i^{(2)}\}$. The different matched pairs $\mathfrak{g}_1 \bowtie \mathfrak{g}_2$ for fixed $\mathfrak{g}_1$ and $\mathfrak{g}_2$, giving rise to inequivalent $N$=2 structures, will then be given by the space of (projectivized) orbits of the admissible $[g_{ij}]$ given in Table 4, under the action of the group basis changes preserving the structure constants of $\mathfrak{g}_1$ and $\mathfrak{g}_2$. This group contains $\operatorname{Aut}\mathfrak{g}_1 \times \operatorname{Aut}\mathfrak{g}_2$ as a subgroup and we will first focus on the orbits of this subgroup. It turns out that this already narrows down the list in Table 4 to just a few constructions. We will then investigate whether the extra freedom in changing basis relates some of the elements of this smaller list any further, but only after we have discarded the nonsolvable solutions.

In other words, we fix $\mathfrak{g}_1$ and $\mathfrak{g}_2$ with basis $\{e_i^{(1)}\}$ and $\{e_i^{(2)}\}$, respectively. We let $f_{ij}^{(1)k}$ and $f_{ij}^{(2)k}$ denote the respective structure constants. Let $\mathcal{M} \subset GL(3, \mathbb{C})$ denote the space of allowed $g_{ij}$ and let $P\mathcal{M} \subset PGL(3, \mathbb{C})$ denote its projectivization. If $A \in \operatorname{Aut}\mathfrak{g}_1 \subset GL(3, \mathbb{C})$ and $B \in \operatorname{Aut}\mathfrak{g}_2 \subset GL(3, \mathbb{C})$, then their action on $P\mathcal{M}$, induced by their action on $\mathcal{M}$, is given by

$$[g] \mapsto [A^t \cdot g \cdot B] \ , \tag{6.1}$$

where · stands for matrix multiplication in $GL(3, \mathbb{C})$, and the brackets around a matrix denote its projective class. Thus two elements $[g_{ij}], [g_{ij}'] \in PGL(3, \mathbb{C})$, lie in the same orbit of $\operatorname{Aut}\mathfrak{g}_1 \times \operatorname{Aut}\mathfrak{g}_2$ if and only if there exist automorphisms $A$ and $B$ of $\mathfrak{g}_1$ and $\mathfrak{g}_2$ respectively and a nonzero complex number $\lambda \in \mathbb{C}^\times$, such that

$$\lambda g' = A^t \cdot g \cdot B \ . \tag{6.2}$$

It is then a straightforward exercise to go through all the pairs $(\mathfrak{g}_1, \mathfrak{g}_2)$ in Table 4, and determine the orbits. The results are summarized in Table 5 below, which also contains information on the structure of the Lie algebra.

|   | Matched Pair | $g_{ij}$ | Structure |
|---|---|---|---|
| I | $\mathfrak{a}_3 \bowtie \mathfrak{a}_3$ | $\begin{pmatrix} 1 & 0 & 0 \\ 0 & 1 & 0 \\ 0 & 0 & 1 \end{pmatrix}$ | Abelian |
| II | $\mathfrak{a}_3 \bowtie \mathfrak{s}_{3,1}$ | $\begin{pmatrix} 1 & 0 & 0 \\ 0 & 1 & 0 \\ 0 & 0 & 1 \end{pmatrix}$ | Nilpotent |
| III | $\mathfrak{a}_3 \bowtie \mathfrak{s}_{3,2}$ | $\begin{pmatrix} 1 & 0 & 0 \\ 0 & 1 & 0 \\ 0 & 0 & 1 \end{pmatrix}$ | Solvable |
| IV($\mu$) | $\mathfrak{a}_3 \bowtie \mathfrak{s}_{3,3}(\mu)$ | $\begin{pmatrix} 1 & 0 & 0 \\ 0 & 1 & 0 \\ 0 & 0 & 1 \end{pmatrix}$ | Solvable |
| V | $\mathfrak{s}_{3,1} \bowtie \mathfrak{s}_{3,1}$ | $\begin{pmatrix} 0 & 0 & 1 \\ 0 & 1 & 0 \\ 1 & 0 & 0 \end{pmatrix}$ | Nilpotent |
| VI$_1$ | $\mathfrak{s}_{3,1} \bowtie \mathfrak{s}_{3,2}$ | $\begin{pmatrix} 0 & 0 & 1 \\ 0 & 1 & 0 \\ 1 & 0 & 0 \end{pmatrix}$ | Solvable |
| VI$_2$ | $\mathfrak{s}_{3,1} \bowtie \mathfrak{s}_{3,2}$ | $\begin{pmatrix} 0 & 1 & 0 \\ 1 & 0 & 0 \\ 0 & 0 & 1 \end{pmatrix}$ | $s\ell(2) \ltimes s\ell(2)^*$ |
| VII$_1$ | $\mathfrak{s}_{3,1} \bowtie \mathfrak{s}_{3,3}(1)$ | $\begin{pmatrix} 0 & 0 & 1 \\ 0 & 1 & 0 \\ 1 & 0 & 0 \end{pmatrix}$ | Solvable |
| VII$_2$ | $\mathfrak{s}_{3,1} \bowtie \mathfrak{s}_{3,3}(1)$ | $\begin{pmatrix} 1 & 0 & 0 \\ 0 & 1 & 0 \\ 0 & 0 & 1 \end{pmatrix}$ | $s\ell(2) \ltimes s\ell(2)^*$ |
| VIII($\mu \neq 1$) | $\mathfrak{s}_{3,1} \bowtie \mathfrak{s}_{3,3}(\mu)$ | $\begin{pmatrix} 0 & 0 & 1 \\ 0 & 1 & 0 \\ 1 & 0 & 0 \end{pmatrix}$ | Solvable |
| IX$_1$ | $\mathfrak{s}_{3,2} \bowtie \mathfrak{s}_{3,3}(-1)$ | $\begin{pmatrix} 0 & 0 & 1 \\ 0 & 1 & 0 \\ 1 & 0 & 0 \end{pmatrix}$ | Solvable |
| IX$_2$ | $\mathfrak{s}_{3,2} \bowtie \mathfrak{s}_{3,3}(-1)$ | $\begin{pmatrix} 0 & 0 & 1 \\ 1 & 1 & 0 \\ 1 & 0 & 0 \end{pmatrix}$ | $s\ell(2) \ltimes s\ell(2)^*$ |
| X | $\mathfrak{s}_{3,3}(1) \bowtie \mathfrak{s}_{3,3}(-1)$ | $\begin{pmatrix} 1 & 0 & 0 \\ 0 & 1 & 0 \\ 0 & 0 & 1 \end{pmatrix}$ | $s\ell(2) \times s\ell(2)$ |
| XI$_1$($\mu \neq \pm 1$) | $\mathfrak{s}_{3,3}(\mu) \bowtie \mathfrak{s}_{3,3}(-\mu)$ | $\begin{pmatrix} 1 & 0 & 0 \\ 0 & 1 & 0 \\ 0 & 0 & 1 \end{pmatrix}$ | $\begin{cases} s\ell(2) \times s\ell(2) \text{ for } \mu \neq 0 \\ s\ell(2) \times \mathfrak{a}_3 \text{ for } \mu = 0 \end{cases}$ |
| XI$_2$($\mu \neq \pm 1$) | $\mathfrak{s}_{3,3}(\mu) \bowtie \mathfrak{s}_{3,3}(-\mu)$ | $\begin{pmatrix} 0 & 0 & 1 \\ 0 & 1 & 0 \\ 1 & 0 & 0 \end{pmatrix}$ | Solvable |
| XI$_3$($\mu \neq \pm 1$) | $\mathfrak{s}_{3,3}(\mu) \bowtie \mathfrak{s}_{3,3}(-\mu)$ | $\begin{pmatrix} 1 & 0 & 0 \\ 0 & 0 & 1 \\ 0 & 1 & 0 \end{pmatrix}$ | Solvable |

**Table 5**  *Equivalence classes of matched pairs*

### The details behind Table 5

The idea behind Table 5 is the following. We take one by one the matched pairs in Table 4 and we study equation (6.2) for each of them using the explicit form of the automorphisms found in Section 4 and displayed in Table 3. We find it easier in fact, to turn equation (6.2) around and study instead the equivalent equation:

$$A^t \cdot g = \lambda g' \cdot B^{-1} \ . \tag{6.3}$$

Then we choose $g'$ to be some reference fixed matrix and study all those matrices $g$ which can be related to $g'$ by equation (6.3). If not all metrics $g$ are in the orbit of $g'$, then we choose another $g'$ for the remaining matrices, and so on until we have swept out all the space of the admissible matrices. Cases I-IV are trivial, since the automorphism group of the abelian algebra $\mathfrak{a}_3$ is the full general linear group, hence given any $g$ and $g'$, (6.3) can be solved for some $A$, regardless what $\lambda$ and $B$ are.

- Case V: $\mathfrak{s}_{3,1} \bowtie \mathfrak{s}_{3,1}$.

We investigate (6.3) where $A$ and $B^{-1}$ are automorphisms of $\mathfrak{s}_{3,1}$. From Table 3 we read off their form:

$$A^t = \begin{pmatrix} \Delta & 0 & 0 \\ v_1 & a_{11} & a_{12} \\ v_2 & a_{12} & a_{22} \end{pmatrix} \quad \text{and} \quad B^{-1} = \begin{pmatrix} \Delta' & v'_1 & v'_2 \\ 0 & a'_{11} & a'_{12} \\ 0 & a'_{12} & a'_{22} \end{pmatrix}, \quad (6.4)$$

where $\Delta = a_{11}a_{22} - a_{12}a_{21}$ and $\Delta' = a'_{11}a'_{22} - a'_{12}a'_{21}$. We try first of all a reference matrix $g'$ of the form:

$$g' = \begin{pmatrix} 0 & 0 & 1 \\ 0 & 1 & 0 \\ 1 & 0 & 0 \end{pmatrix}. \quad (6.5)$$

Then the question is whether all admissible $g$'s—in this case, all nondegenerate $g$'s with $g_{11} = 0$—can be chosen to satisfy (6.3) for some $A$ and $B$ given above, with the proviso that $\Delta$, $\Delta'$ and $\lambda$ are different from zero. Plugging (6.4) and (6.5) into (6.3) we get eight equations – six of which simply define what $v'_i$ and $a'_{ij}$ are in terms of the other variables. In particular we have

$$\begin{aligned} \lambda a'_{11} &= v_1 g_{12} + a_{11} g_{22} + a_{12} g_{32} \\ \lambda a'_{12} &= v_1 g_{13} + a_{11} g_{23} + a_{12} g_{33} \\ \lambda a'_{21} &= \Delta g_{12} \\ \lambda a'_{22} &= \Delta g_{13}, \end{aligned} \quad (6.6)$$

From where we read off what $\Delta'$ should be:

$$\lambda^2 \Delta' = \Delta \left( a_{11}(g_{13}g_{22} - g_{12}g_{23}) + a_{12}(g_{13}g_{32} - g_{12}g_{33}) \right). \quad (6.7)$$

On the other hand, the other two equations from (6.3) say that

$$\begin{aligned} \lambda \Delta' &= a_{21} g_{21} + a_{22} g_{31} \\ 0 &= a_{11} g_{21} + a_{12} g_{31}. \end{aligned} \quad (6.8)$$

Now notice that because $g_{11} = 0$ but $\det g \neq 0$, then $g_{21}$ and $g_{31}$ cannot be both zero. Assume first that $g_{21} \neq 0$. Then we can use the previous equation to solve for $a_{11}$:

$$a_{11} = -\frac{g_{31}}{g_{21}} a_{12}, \quad (6.9)$$

which into (6.7) yields

$$\lambda \Delta' = \Delta \frac{a_{12}}{g_{21}} \det g. \quad (6.10)$$

It follows from this that we can choose $a_{12}$, $a_{21}$ and $a_{22}$ subject to the first equation in (6.8) in such a way that both $\Delta$ and $\Delta'$ are nonzero. A similar argument works in the case that $g_{21} = 0$, since then $g_{31} \neq 0$. Thus we conclude that all admissible $g_{ij}$ lie in the orbit of the $g'$ in (6.5). Moreover notice that since we did not have to fix the value of $\lambda$, it follows that this is true even before projectivizing.

• **Case VI:** $\mathfrak{s}_{3,1} \bowtie \mathfrak{s}_{3,2}$.

Since $g_{11} = 0$ let us first try $g'$ given again by (6.5) above. Equation (6.3) now becomes:

$$\lambda \begin{pmatrix} 0 & 0 & 1 \\ 0 & 1 & 0 \\ 1 & 0 & 0 \end{pmatrix} \begin{pmatrix} a' & b' & v'_1 \\ 0 & a' & v'_2 \\ 0 & 0 & 1 \end{pmatrix} = \begin{pmatrix} \Delta & 0 & 0 \\ v_1 & a_{11} & a_{12} \\ v_2 & a_{12} & a_{22} \end{pmatrix} \begin{pmatrix} 0 & g_{12} & g_{13} \\ g_{21} & g_{22} & g_{33} \\ g_{31} & g_{32} & g_{33} \end{pmatrix}, \quad (6.11)$$

from where it follows that $g_{12} = 0$. Therefore not all $(g_{ij})$ will lie in the same orbit this time.

◦ **Subcase** $\text{VI}_1$: $g_{12} = 0$.

In this case $g_{13} \neq 0$, whence the equation $\lambda = \Delta g_{13}$ is consistent. The only relevant equations from (6.3) are in this case

$$\begin{aligned} \lambda a' &= a_{21} g_{21} + a_{22} g_{31} \\ \lambda a' &= a_{11} g_{22} + a_{12} g_{32} \\ 0 &= a_{11} g_{21} + a_{12} g_{31}. \end{aligned} \quad (6.12)$$

Since $g_{11} = g_{12} = 0$, the submatrix

$$\begin{pmatrix} g_{21} & g_{22} \\ g_{31} & g_{32} \end{pmatrix} \quad (6.13)$$

is nonsingular, hence the last two equations in (6.12) can be solved for $a_{11}$ and $a_{12}$ and for any nonzero $a'$. To solve the first equation in (6.12), simply note that $g_{21}$ and $g_{31}$ cannot be zero. Notice moreover that $a'$ remains a free parameter which we can choose at will. This freedom translates in the freedom to rescale $\Delta$, hence $\lambda$ is free. In other words, all admissible $g_{ij}$ with in addition $g_{12} = 0$ are in the orbit of $g'$ in (6.5) even before projectivizing.

◦ **Subcase** $\text{VI}_2$: $g_{12} \neq 0$.

We try then the following:

$$g' = \begin{pmatrix} 0 & 1 & 0 \\ 1 & 0 & 0 \\ 0 & 0 & 1 \end{pmatrix} . \qquad (6.14)$$

In this case, the relevant equations coming out from (6.3) are

$$\begin{aligned} \lambda a' &= \Delta g_{12} \\ \lambda a' &= a_{11} g_{21} + a_{12} g_{31} \\ 0 &= a_{21} g_{21} + a_{22} g_{31} \\ \lambda &= v_2 g_{13} + a_{21} g_{23} + a_{22} g_{33} . \end{aligned} \qquad (6.15)$$

The first one determines $a'$. The second and third can be solved for the $a_{ij}$ because $g_{21}$ and $g_{31}$ cannot both be zero. The freedom left in the $a_{ij}$ is enough to be able to be able to solve the last equation for any value of $\lambda$, even when $g_{13} = 0$. Hence all admissible $g_{ij}$ with $g_{12} \neq 0$ are in the orbit of (6.14) even before projectivization.

• Case VII: $\mathfrak{s}_{3,1} \bowtie \mathfrak{s}_{3,3}(1)$.

Since any $g$ is admissible, we first try

$$g' = \begin{pmatrix} 1 & 0 & 0 \\ 0 & 1 & 0 \\ 0 & 0 & 1 \end{pmatrix} . \qquad (6.16)$$

Equation (6.3) now becomes:

$$\lambda \begin{pmatrix} a'_{11} & a'_{12} & v'_1 \\ a'_{21} & a'_{22} & v'_2 \\ 0 & 0 & 1 \end{pmatrix} = \begin{pmatrix} \Delta & 0 & 0 \\ v_1 & a_{11} & a_{12} \\ v_2 & a_{12} & a_{22} \end{pmatrix} \begin{pmatrix} g_{11} & g_{12} & g_{13} \\ g_{21} & g_{22} & g_{33} \\ g_{31} & g_{32} & g_{33} \end{pmatrix} . \qquad (6.17)$$

Of the nine equations contained above, six of them define $a'_{ij}$ and $v'_i$, and the other three can be written as

$$g^t \cdot \begin{pmatrix} v_2 \\ a_{21} \\ a_{22} \end{pmatrix} = \begin{pmatrix} 0 \\ 0 \\ \lambda \end{pmatrix} \qquad (6.18)$$

which can be uniquely solved for $v_2$ and $a_{2i}$ for any $\lambda$, due to the fact that $g$ is invertible. Notice that $a_{1i}$ can be moreover chosen at will in such a way that $\Delta$ is nonzero. Hence the only constraint is to make sure that $(a'_{ij})$ is invertible. The entries of $(a'_{ij})$ can be read off from (6.17):

$$\begin{aligned} \lambda a'_{11} &= \Delta g_{11} \\ \lambda a'_{12} &= \Delta g_{12} \\ \lambda a'_{21} &= v_1 g_{11} + a_{11} g_{21} + a_{12} g_{31} \\ \lambda a'_{22} &= v_1 g_{12} + a_{11} g_{22} + a_{12} g_{32} . \end{aligned} \qquad (6.19)$$

Computing its determinant, we find

$$\Delta^{-1} \lambda^2 \, \|a'_{ij}\| = a_{11} \begin{Vmatrix} g_{11} & g_{12} \\ g_{21} & g_{22} \end{Vmatrix} + a_{12} \begin{Vmatrix} g_{11} & g_{12} \\ g_{31} & g_{32} \end{Vmatrix} , \qquad (6.20)$$

which can be chosen to be nonzero provided that not both of the above minors of $(g_{ij})$ are zero. How about if they are? If this is the case then it must happen that $g_{11} = g_{12} = 0$. Thus we have two subcases.

∘ Subcase VII$_1$ : $g_{11} \neq 0$ or $g_{12} \neq 0$.

All these $g_{ij}$ lie in the orbit of $g'$ in (6.16) without the need to projectivize.

∘ Subcase VII$_2$ : $g_{11} = g_{12} = 0$.

In this case, $g_{13} \neq 0$ and we try $g'$ given again by (6.5). There are seven nontrivial equations to (6.3) now, two of which simply define $v'_i$. The other five can be written as $\lambda = \Delta g_{13}$ which relates $\Delta$ and $\lambda$; and

$$\begin{pmatrix} a_{11} & a_{12} \\ a_{21} & a_{22} \end{pmatrix} \begin{pmatrix} g_{21} & g_{22} \\ g_{31} & g_{32} \end{pmatrix} = \begin{pmatrix} a'_{21} & a'_{22} \\ a'_{11} & a'_{12} \end{pmatrix} . \qquad (6.21)$$

But notice that the submatrix $\begin{pmatrix} g_{21} & g_{22} \\ g_{31} & g_{32} \end{pmatrix}$ is invertible, so that this always has solution for $(a'_{ij})$ regardless what $(a_{ij})$ is. Notice also that $\Delta$ and hence $\lambda$ remain free; whence all the admissible pairings $(g_{ij})$ with $g_{11} = g_{12} = 0$ lie in the orbit of $g'$ in (6.5) without having to projectivize.

• Case VIII($\mu \neq 1$): $\mathfrak{s}_{3,1} \bowtie \mathfrak{s}_{3,3}(\mu \neq 1)$.

Since the admissible $g_{ij}$ must have $g_{11} = g_{12} = 0$, then we try $g'$ given by (6.5). In this case, equation (6.3) becomes

$$\lambda \begin{pmatrix} 0 & 0 & 1 \\ 0 & b' & v'_2 \\ a' & 0 & v'_1 \end{pmatrix} = \begin{pmatrix} \Delta & 0 & 0 \\ v_1 & a_{11} & a_{12} \\ v_2 & a_{12} & a_{22} \end{pmatrix} \begin{pmatrix} 0 & 0 & g_{13} \\ g_{21} & g_{22} & g_{33} \\ g_{31} & g_{32} & g_{33} \end{pmatrix} . \qquad (6.22)$$

There are seven equations here of which two simply define $v'_i$ and one relates $\lambda$ and $\Delta$: $\lambda = \Delta g_{13}$, which is consistent since $g_{13} \neq 0$. The remaining four equations can be written as:

$$\lambda \begin{pmatrix} 0 & b' \\ a' & 0 \end{pmatrix} = \begin{pmatrix} a_{11} & a_{12} \\ a_{12} & a_{22} \end{pmatrix} \begin{pmatrix} g_{21} & g_{22} \\ g_{31} & g_{32} \end{pmatrix} . \quad (6.23)$$

Since the $g$ submatrix is nondegenerate, this can be solved for $a'$ and $b'$ regardless what $(a_{ij})$ are. In particular, we are free to choose $\Delta$, and hence $\lambda$. In other words, all admissible pairings lie in the orbit of $g'$ without projectivizing.

- Case IX: $\mathfrak{s}_{3,2} \bowtie \mathfrak{s}_{3,3}(-1)$.

The admissible $g_{ij}$ have $g_{11} = g_{12} = 0$ in this case; so we try $g'$ as in (6.5) again. Equation (6.3) turns out to be now:

$$\lambda \begin{pmatrix} 0 & 0 & 1 \\ 0 & b' & v_2' \\ a' & 0 & v_1' \end{pmatrix} = \begin{pmatrix} a & 0 & 0 \\ b & a & 0 \\ c & d & 1 \end{pmatrix} \begin{pmatrix} 0 & 0 & g_{13} \\ g_{21} & g_{22} & g_{33} \\ g_{31} & g_{32} & g_{33} \end{pmatrix} , \quad (6.24)$$

or

$$\lambda \begin{pmatrix} 0 & 0 & -1 \\ b' & 0 & v_2' \\ 0 & a' & v_1' \end{pmatrix} = \begin{pmatrix} a & 0 & 0 \\ b & a & 0 \\ c & d & 1 \end{pmatrix} \begin{pmatrix} 0 & 0 & g_{13} \\ g_{21} & g_{22} & g_{33} \\ g_{31} & g_{32} & g_{33} \end{pmatrix} , \quad (6.25)$$

depending on the form of the automorphism of $\mathfrak{s}_{3,3}(-1)$ that we take. In either case, it is clear that this cannot be satisfied by all $g_{ij} = 0$. The following subcases present themselves.

○ Subcase IX$_1$ : $g_{21}g_{22} = 0$.

If this is the case, then we know that $g_{21}$ and $g_{22}$ cannot both be zero. If $g_{21} = 0$ then we use (6.24) and if $g_{22} = 0$ we use (6.25). It either case it is easy to show that the relevant equation can be solved for any $a$, hence for any $\lambda$. Thus all admissible pairings with $g_{21}g_{22} = 0$ lie in the orbit of (6.5) without having to projectivize.

○ Subcase IX$_2$ : $g_{21}g_{22} \neq 0$.

This case suggests that we try a more complicated reference pairing:

$$g' = \begin{pmatrix} 0 & 0 & 1 \\ 1 & 1 & 0 \\ 0 & 1 & 0 \end{pmatrix} , \quad (6.26)$$

which turns (6.24) into the following:

$$\lambda \begin{pmatrix} 0 & 0 & 1 \\ a' & b' & v_1' + v_2' \\ 0 & b' & v_2' \end{pmatrix} = \begin{pmatrix} a & 0 & 0 \\ b & a & 0 \\ c & d & 1 \end{pmatrix} \begin{pmatrix} 0 & 0 & g_{13} \\ g_{21} & g_{22} & g_{33} \\ g_{31} & g_{32} & g_{33} \end{pmatrix} . \quad (6.27)$$

In this case, there is not sufficient freedom to leave $\lambda$ unspecified. Indeed, of the seven equations in (6.27), two of them serve to define $v_i'$. Four of the remaining five equations fix $d$, $a'$, $b'$ and the ratio $a/\lambda$; but the fifth equation fixes $\lambda$ to be given by

$$\lambda = \frac{\|g_{ij}\|}{g_{21}g_{22}} . \quad (6.28)$$

In other words, every admissible $g_{ij}$ is conformal to one in the orbit of (6.26).

- Case X: $\mathfrak{s}_{3,3}(1) \bowtie \mathfrak{s}_{3,3}(-1)$.

Since any nonsingular $(g_{ij})$ will do we first of all try $g'$ given by (6.16). In this case, (6.3) becomes:

$$\lambda \begin{pmatrix} a' & 0 & v_1' \\ 0 & b' & v_2' \\ 0 & 0 & 1 \end{pmatrix} = \begin{pmatrix} a_{11} & a_{12} & 0 \\ a_{21} & a_{22} & 0 \\ v_1 & v_2 & 1 \end{pmatrix} \begin{pmatrix} g_{11} & g_{12} & g_{13} \\ g_{21} & g_{22} & g_{33} \\ g_{31} & g_{32} & g_{33} \end{pmatrix} . \quad (6.29)$$

It is in fact easier to analyze by turning around and considering the most general metric that lies in the orbit of $g'$. It is clearly seen to be given by the product of matrices of the form

$$\lambda \begin{pmatrix} A & 0 \\ v^t & 1 \end{pmatrix} \begin{pmatrix} D & v' \\ 0 & 1 \end{pmatrix} , \quad (6.30)$$

where $A, D \in GL(2, \mathbb{C})$ with $D$ is diagonal, and $v, v' \in \mathbb{C}^2$. It is clear that any nonsigular $g$ can be written as such a product, but maybe needing to fix $\lambda$. In other words, any admissible pairing $(g_{ij})$ is conformal to one in the orbit of $g'$.

- Case XI($\mu \neq \pm 1$): $\mathfrak{s}_{3,3}(\mu \neq \pm 1) \bowtie \mathfrak{s}_{3,3}(-\mu)$.

In the final case, the admissible pairings have $g_{12} = g_{21} = 0$. We can therefore first try $g'$ given by (6.16). Equation (6.3) now becomes:

$$\lambda \begin{pmatrix} a' & 0 & v_1' \\ 0 & b' & v_2' \\ 0 & 0 & 1 \end{pmatrix} = \begin{pmatrix} a & 0 & 0 \\ 0 & b & 0 \\ v_1 & v_2 & 1 \end{pmatrix} \begin{pmatrix} g_{11} & 0 & g_{13} \\ 0 & g_{22} & g_{33} \\ g_{31} & g_{32} & g_{33} \end{pmatrix} . (6.29)$$

Clearly not all pairings $(g_{ij})$ satisfy this equation. The following subcases arise.

○ Subcase XI$_1$($\mu$): $g_{11}g_{22} \neq 0$.

If this is the case, then (6.29) can be solved for any such pairing, provided we fix $\lambda$ to be

$$\lambda = -\frac{g_{31}g_{13}}{g_{11}} - \frac{g_{32}g_{23}}{g_{22}} + g_{33} = \frac{\|g_{ij}\|}{g_{11}g_{22}} . \quad (6.31)$$

Nondegeracy of the pairing forbids $g_{11}$ and $g_{22}$ from vanishing simultaneously. This gives rise to the two last subcases.

∘ Subcase $XI_2(\mu)$: $g_{11} = 0$.

For this case we try $g'$ to be of the form (6.5). Working through (6.3) in this case, shows that all of the resulting six equations can be solved without having to fix $\lambda$. Hence all admissible pairings with $g_{11} = 0$ lie in the orbit of (6.5) without having to projectivize.

∘ Subcase $XI_3(\mu)$: $g_{22} = 0$.

This case is identical to the previous one, except that we take the reference pairing to be

$$g' = \begin{pmatrix} 1 & 0 & 0 \\ 0 & 0 & 1 \\ 0 & 1 & 0 \end{pmatrix} . \qquad (6.32)$$

In particular there is no need to projectivize, since $\lambda$ is not fixed.

Discarding solutions which are not solvable

Notice that it is not necessary for $\mathfrak{g}_1 \bowtie \mathfrak{g}_2$ to be solvable even if $\mathfrak{g}_1$ and $\mathfrak{g}_2$ are. Thus we must check which of the Lie algebras $\mathfrak{g}_1 \bowtie \mathfrak{g}_2$ in Table 5 are solvable. The Lie brackets for the six-dimensional Lie algebras $\mathfrak{g}_1 \bowtie \mathfrak{g}_2$ in Table 5 are easy enough to write down with the information provided, and, in principle, one could compute the series of derived ideals and checks whether it terminates. However, knowing that $\mathfrak{g}_1 \bowtie \mathfrak{g}_2$ are self-dual, there exists an easier test for nonsolvability, which will allow us to rule out many of the examples at once. If $\mathfrak{g}$ is a solvable self-dual Lie algebra, then it must have a nontrivial center. This can be seen in two ways. First of all, notice that if $\mathfrak{g}$ is self-dual, then the orthogonal complement of its first derived ideal $[\mathfrak{g}, \mathfrak{g}]$ agrees with the center $\mathfrak{z}$ of $\mathfrak{g}$: $x \in \mathfrak{z} \Leftrightarrow \langle [x, \mathfrak{g}], \mathfrak{g} \rangle = 0 \Leftrightarrow \langle x, [\mathfrak{g}, \mathfrak{g}] \rangle = 0$. Alternatively, and in terms of our discussion of double extensions, notice that if $\mathfrak{h}$ is abelian (in particular, one-dimensional) then $\mathfrak{h}^*$ is actually central in the double extension $\mathfrak{D}(\mathfrak{g}, \mathfrak{h})$. It is much easier to decide which of the algebras in Table 5 have indeed a nontrivial center. A relatively short calculation reveals that the following Lie algebras have *no* center: $VI_2$, $VII_2$, $IX_2$, $X$, and $XI_1(\mu \neq 0, \pm 1)$. Therefore we can safely discard them from our classification since they are not solvable, indeed it is easy to see that $VI_2 \cong VII_2 \cong IX_2 \cong s\ell(2) \ltimes s\ell(2)^*$, and $X \cong XI_1(\mu \neq 0) \cong s\ell(2) \times s\ell(2)$. All the remaining algebras except for one: $XI_1(0) \cong s\ell(2) \times \mathfrak{a}_3$, turn out to be solvable. We will see this explicitly in the next section when we exhibit the structure of these self-dual Lie algebras in terms of double extensions.

There is one final point that remains addressing. When determining the orbits of the space $P\mathcal{M}$ of conformal classes of allowed pairings $g_{ij}$ under basis changes preserving $\mathfrak{g}_1$ and $\mathfrak{g}_2$, we restricted ourselves to basis changes belonging to the subgroup $\mathrm{Aut}\, \mathfrak{g}_1 \times \mathrm{Aut}\, \mathfrak{g}_2$. We should check whether conformal classes of pairings which belong to different orbits of this subgroup can in fact lie in the same orbit of the full group of basis transformations. Among the solvable Lie algebras in Table 5, we see that the only possibility of this happening is for the pair $XI_{2,3}(\mu \neq \pm 1)$. We will see in the next section that in this case, both Lie algebras as isomorphic as self-dual Lie algebras; that is, they are isomorphic as Lie algebras, and the isomorphism preserves the invariant metric. And this implies that they lead to isomorphic $N$=2 structures, since the isomorphism restricts to an isomorphism of the three-dimensional Lie algebras forming the matched pair. An explicit automorphism is given by

$$\begin{aligned} XI_2(\mu) &\longrightarrow XI_3(\mu) \\ e_1 &\mapsto \tfrac{1}{2}(e_2 - e_4 - e_5) \\ e_2 &\mapsto \tfrac{1}{2}(e_3 + e_6) \\ e_3 &\mapsto e_1 - \tfrac{1}{2}(e_3 - e_6) \\ e_4 &\mapsto \tfrac{1}{2}(e_2 + e_4 - e_5) \\ e_5 &\mapsto e_2 + e_5 \\ e_6 &\mapsto -e_1 - \tfrac{1}{2}(e_3 - e_6) \end{aligned} \qquad (6.33)$$

In summary, the isomorphism classes of $c$=9 $N$=2 structures on solvable Lie algebras are (in the notation of Table 5) the following:

$$\boxed{\text{I-V}, \ VI_1, \ VII_1, \ VIII(\mu \neq 1), \ IX_1, \ \text{and } XI_2(\mu \neq \pm 1) \ .} \qquad (6.34)$$

§7   Understanding the solutions as self-dual Lie algebras

The solvable Lie algebras found in the previous section are particular examples of self-dual Lie algebras, for which a structure theorem exists. In this section we would like to use this structure theorem to make their Lie algebraic structure more transparent. This will facilitate their comparison and will allow us to prove, in particular, that those Lie algebras which we not ruled out by testing for a center, are inded solvable as we claimed.

The structure theorem of [**21**] actually gives a reasonably constructive way to write down an indecomposable self-dual Lie algebra as a double extension. Suppose that $\mathfrak{d}$ is an indecomposable self-dual Lie algebra and let $I$ be a minimal ideal; that is, an ideal of $\mathfrak{d}$ not properly containing any other ideal. Clearly $I^\perp$ is a maximal ideal and hence $I \subset I^\perp$. Indeed, $I \cap I^\perp$ is an ideal contained in $I$, hence it has to be either zero or all of $I$; but if it were zero, then $\mathfrak{d} =$ $I \times I^\perp$, violating indecomposability. Let $\mathfrak{g} \equiv I^\perp/I$. It is clearly a self-dual Lie algebra. Since $I^\perp$ is a maximal ideal, $\mathfrak{h} \equiv \mathfrak{d}/I^\perp$ cannot have any proper ideals: hence it is either simple or one-dimensional. Because of this, it also follows that $\mathfrak{h}$ is actually a subalgebra of $\mathfrak{d}$ so that $\mathfrak{d}$ is semidirect product (that is, a split extension) of $I^\perp$ by $\mathfrak{h}$; whereas $I^\perp$ itself is a central extension of $\mathfrak{g}$ by $I$. Moreover it follows that $\mathfrak{h}$ and $I$ are nondegenerately (and equivariantly) paired so that $I \cong \mathfrak{h}^*$. In summary, $\mathfrak{d} \cong \mathfrak{D}(\mathfrak{g}, \mathfrak{h})$.

Using the above "algorithm" it is possible to go one by one through the matched pairs in Table 5 and decompose them in terms of direct sums and double extensions. We will see that except for $\mathfrak{a}_3 \bowtie \mathfrak{a}_3 \cong \mathfrak{a}_6$ (Case I), they are all double extensions of the abelian algebra $\mathfrak{a}_4$ by $\mathfrak{a}_1$. The results are summarized in Table 6. The notation may need some explanation. As we saw in Section 4, a double extension $\mathfrak{D}_\rho(\mathfrak{g}, \mathfrak{a}_1)$ is characterized uniquely by the Lie brackets and metric of $\mathfrak{g}$ and the matrix defining the action of $\mathfrak{a}_1$ on $\mathfrak{g}$. In the algebras in Table 6, $\mathfrak{g}$ is always abelian. We then fix a basis $\{x_1, x_2, x_3, x_4\}$ for $\mathfrak{g}$ in such a way that the nonzero inner products are given by $\langle x_1, x_2 \rangle = \langle x_3, x_4 \rangle = 1$. Then in Table 6 we simply list the matrices $\rho_i{}^j$ corresponding to the action of $h$ on $\mathfrak{g}$. The brackets and the metric for the double extension then follow from (4.6) and (4.7).

Except for the abelian case, the only other previously known example is IV($\mu = 1$), which corresponds to the Heisenberg algebra discussed by Mohammedi in [18]. Notice that provided that we set $\mu$ to a real number, all of these examples admit a real form.

In the above form, it is immediately clear that many of these Lie algebras are isomorphic as self-dual Lie algebras; although, this isomorphism does not extend to an isomorphism of $N$=2 structures. From Table 6, we can read the isomorphism classes of self-dual Lie algebras which admit an $N$=2 structure. Since the self-dual Lie algebras above are all double extensions of the form $\mathfrak{D}_\rho(\mathfrak{g}, \mathfrak{a}_1)$, and they are characterized uniquely by the matrix of inner products on $\mathfrak{g}$ and by the matrix $\rho$ defining the action of $\mathfrak{a}_1$ on $\mathfrak{g}$; we see that two such double extensions will be isomorphic as self-dual Lie algebras, if and only if there exists a vector space isomorphism $\phi: \mathfrak{g} \to \mathfrak{g}$, which preserves the metric and which intertwines between the $\mathfrak{a}_1$-actions. Having chosen a basis for $\mathfrak{g}$, we can write these conditions in terms of matrices in $\mathrm{Mat}(4, \mathbb{C})$. We then have that two double extensions with data $(\rho, g)$ and $(\rho', g')$, where we understand each pair as a pair of matrices in $\mathrm{Mat}(4, \mathbb{C})$, define isomorphic self-dual Lie algebras, if and only if there exists a matrix $\phi \in GL(4, \mathbb{C})$ such that

$$\phi^t \cdot g \cdot \phi = g' \quad \text{and} \quad \rho' \cdot \phi = \varepsilon \phi \cdot \rho \;, \tag{7.1}$$

where $\cdot$ stands for matrix multiplication and $\varepsilon$ is a nonzero scale—that is, an automorphism of $\mathfrak{a}_1$. In our case, we have chosen bases in such a way that $g = g'$. By inspection we already see the following isomorphisms of self-dual Lie algebras: IV($\mu$) $\cong$ XI$_{2,3}$($\mu$), for all $\mu$ such that they are all defined. Furthermore a little computation shows also that III $\cong$ VI$_1$ $\cong$ IX$_1$, whereas VII$_1$ $\cong$ IV(1) and VII$_1$ $\not\cong$ III. Also VIII($\mu$) $\cong$ IV($\mu$) for $\mu \neq 1$; whereas II $\not\cong$ V. Notice moreover that IV($\mu$) $\cong$ IV($-\mu$). Summarising, we have four isomorphism classes of self-dual Lie algebras—one of which with a parameter:

$$A : \text{II};$$
$$B : \text{V};$$
$$C : \text{III, VI}_1, \text{IX}_1;$$
$$D(0) : \text{IV}(0), \text{VIII}(0), \text{XI}_{2,3}(0);$$
$$D(1) : \text{IV}(1), \text{VII}_1;$$
$$D(-1) : \text{IV}(-1), \text{VIII}(-1);$$
$$D(\nu) : \text{IV}(\nu), \text{VIII}(\nu), \text{XI}_{2,3}(\nu);$$
$$D(\mu) : \text{IV}(\mu), \text{VIII}(\mu), \text{XI}_{2,3}(\mu);$$

where $\mu$ takes all complex values except for the origin and the unit circle, and $\nu$ lies on the unit circle in the complex plane excised of $\pm 1$. Finally we have the additional isomorphisms $D(\mu) \cong D(-\mu)$ and $D(\nu) \cong D(1/\nu)$.

| Algebra | $\rho_i{}^j$ | Algebra | $\rho_i{}^j$ |
|---|---|---|---|
| I | $\begin{pmatrix} 0 & 0 & 0 & 0 \\ 0 & 0 & 0 & 0 \\ 0 & 0 & 0 & 0 \\ 0 & 0 & 0 & 0 \end{pmatrix}$ | VI$_1$ | $\begin{pmatrix} 1 & 0 & -1 & 1 \\ 0 & -1 & 0 & 0 \\ 0 & -1 & -1 & 0 \\ 0 & 1 & 0 & 1 \end{pmatrix}$ |
| II | $\begin{pmatrix} 0 & 0 & 0 & 0 \\ 0 & 0 & 1 & 0 \\ 0 & 0 & 0 & 0 \\ -1 & 0 & 0 & 0 \end{pmatrix}$ | VII$_1$ | $\begin{pmatrix} 1 & 0 & -1 & 0 \\ 0 & -1 & 0 & 0 \\ 0 & 0 & -1 & 0 \\ 0 & 1 & 0 & 1 \end{pmatrix}$ |
| III | $\begin{pmatrix} 1 & 0 & 1 & 0 \\ 0 & -1 & 0 & 0 \\ 0 & 0 & 1 & 0 \\ 0 & -1 & 0 & -1 \end{pmatrix}$ | VIII($\mu \neq 1$) | $\begin{pmatrix} 1 & 0 & -1 & 0 \\ 0 & -1 & 0 & 0 \\ 0 & 0 & -\mu & 0 \\ 0 & 1 & 0 & \mu \end{pmatrix}$ |
| IV($\mu$) | $\begin{pmatrix} 1 & 0 & 0 & 0 \\ 0 & -1 & 0 & 0 \\ 0 & 0 & \mu & 0 \\ 0 & 0 & 0 & -\mu \end{pmatrix}$ | IX$_1$ | $\begin{pmatrix} -1 & 0 & -1 & 0 \\ 0 & 1 & 0 & 0 \\ 0 & 0 & -1 & 0 \\ 0 & 1 & 0 & 1 \end{pmatrix}$ |
| V | $\begin{pmatrix} 0 & 0 & 0 & -1 \\ 0 & 0 & 0 & 1 \\ -1 & 1 & 0 & 0 \\ 0 & 0 & 0 & 0 \end{pmatrix}$ | XI$_{2,3}$($\mu \neq \pm 1$) | $\begin{pmatrix} 1 & 0 & 0 & 0 \\ 0 & -1 & 0 & 0 \\ 0 & 0 & \mu & 0 \\ 0 & 0 & 0 & -\mu \end{pmatrix}$ |

**Table 6** *The solvable Lie algebras of Table 5*

We end with some further remarks on the structure of the self-dual Lie algebras discussed above. The matrix $\rho$ which determines the action of $\mathfrak{a}_1$ on $\mathfrak{a}_4$, carries most of the Lie algebraic information on the resulting double extension. In particular it allows us to see if the Lie algebra is nilpotent and it comes into the determination of the Killing form. It is easy to see that the Killing form $\kappa$ for the solvable Lie algebras discussed above, is zero except possibly the entry $\kappa(h, h) = \text{Tr}\,\rho^2$. This is nonzero except for II and V. This implies that the Lie groups corresponding to these Lie algebras are Ricci flat relative to the natural bi-invariant metric; in particular they are Einstein manifolds. Furthermore these algebras are both nilpotent. In fact, it is not hard to show that a double extension $\mathfrak{D}_\rho(\mathfrak{a}, \mathfrak{a}_1)$ where $\mathfrak{a}$ is any abelian algebra and $\rho$ defines the $\mathfrak{a}_1$ action on $\mathfrak{a}$, is nilpotent if and only if the matrix $\rho$ is. In the cases at hand, one has that for II, $\rho^2 = 0$ and for V, $\rho^3 = 0$.

## §8  Conclusions

Let us briefly summarize the highlights of this paper. A real Lie algebra $\mathfrak{g}$ is said to admit an $N{=}2$ structure, if it is self-dual and in addition possesses an integrable complex structure compatible with the metric. We saw in Section 3 that this in turn implies that its complexification $\mathfrak{g}_\mathbb{C}$ admits a vector space decomposition $\mathfrak{g}_\mathbb{C} = \mathfrak{g}_+ \oplus \mathfrak{g}_-$ into isotropic subalgebras; that is, $(\mathfrak{g}_\mathbb{C}, \mathfrak{g}_+, \mathfrak{g}_-)$ is a complex Manin triple. In other words, we have an assignment of a complex Manin triple to every $N{=}2$ structure on a real Lie algebra. This assignment is not invertible, however. That is, not every complex Manin triple $(\mathfrak{m}, \mathfrak{m}_+, \mathfrak{m}_-)$ arises in this fashion, since $\mathfrak{m}$ need not be the complexification of any real Lie algebra. Nevertheless, by the same token, there is an honest bijective correspondence between $N{=}2$ structures on a complex Lie algebra, and complex Manin triples. We therefore set out, as a first step, to classify the $c{=}9$ $N{=}2$ structures on complex solvable Lie algebras.

We restrict to solvable Lie algebras since the problem then becomes finite. It follows easily from the expression of the Virasoro central charge, that for a solvable Lie algebra, the central charge is proportional to the dimension of the algebra—the constant of proportionality being such that to obtain $c{=}9$ we need to consider six-dimensional Lie algebras. We then set out to classify all the complex Manin triples $(\mathfrak{m}, \mathfrak{m}_+, \mathfrak{m}_-)$ with $\dim_\mathbb{C} \mathfrak{m}_\pm = 3$. To this effect it proved convenient to use the concept of a matched pair of Lie algebras: two Lie algebras $\mathfrak{g}_1$ and $\mathfrak{g}_2$ are said to be *matched* if they fit inside a Manin triple $(\mathfrak{g}_1 \bowtie \mathfrak{g}_2, \mathfrak{g}_1, \mathfrak{g}_2)$ where $\mathfrak{g}_1 \bowtie \mathfrak{g}_2$ is some self-dual Lie algebra containing $\mathfrak{g}_1$ and $\mathfrak{g}_2$ as complementary isotropic subalgebras. Demanding that $\mathfrak{g}_1 \bowtie \mathfrak{g}_2$ be solvable, implies that $\mathfrak{g}_1$ and $\mathfrak{g}_2$ are solvable; although the converse need not (and does not) hold. Therefore we first classify the complex solvable three-dimensional Lie algebra, which is achieved in Section 4. This borrows from the classic work of Bianchi, who classified all three-dimensional real Lie algebras: we simply extend the base field and repackage the information in a convenient way. The results are displayed in Table 2. The next step was to see which pairs $(\mathfrak{g}_1, \mathfrak{g}_2)$ of three-dimensional complex solvable Lie algebras can be matched and in how many ways. We did this in Section 5 and the results are in Table 4. In Section 6 we started the analysis of the moduli space of solutions, by partitioning the Manin triples found in Section 5 into isomorphism classes. For the purposes of this discussion we define two Manin triples $(\mathfrak{m}, \mathfrak{m}_+, \mathfrak{m}_-)$ and $(\mathfrak{m}', \mathfrak{m}'_+, \mathfrak{m}'_-)$ to be isomorphic if there exists a Lie algebra isomorphism $\mathfrak{m} \cong \mathfrak{m}'$ which restricts to Lie algebra isomorphisms $\mathfrak{m}_\pm \cong \mathfrak{m}'_\pm$ and which preserves the conformal class of the metric. Strictly speaking, the form of the $N{=}2$ generators will differ unless the isomorphism $\mathfrak{m} \cong \mathfrak{m}'$ is also an isometry, but all essential properties of the $N{=}2$ CFT (for example, the chiral ring) remain the same under a conformal isomorphism. Furthermore, this weaker notion of equivalence does not represent any real loss in generality for the solvable case. The different isomoprhism classes are displayed in Table 5. After having narrowed down the list in Section 5 considerably, we discard those matched pairs which are not solvable. The remaining list of solutions are displayed in (6.34). From this results the final classification of isomorphism classes (in the generalized sense explained above) of $c{=}9$ $N{=}2$ structures on complex solvable Lie algebras. Finally in Section 7 we interpreted these solutions as self-dual Lie algebras.

We conclude now with a discussion of reality. Suppose we want to classify the $c{=}9$ $N{=}2$ structures on *real* a solvable Lie algebra $\mathfrak{g}$. Then the complexification $\mathfrak{g}_\mathbb{C}$ of $\mathfrak{g}$ will be isomorphic over $\mathbb{C}$ to one of the solutions in the list in Section 6. Does that mean that the real classification will be a subset of the complex classification? The answer is of course negative. The reason is that a complex Lie algebra may have more than one nonisomorphic real form, a familiar fact from the classification of simple Lie algebras. Thus in order to obtain a "real" classification, we must still take the isomorphism classes of complex solutions found in Section 6 and classify all the compatible real structures; where the compatibility criterion is that the conjugation $\sigma$ defining the real structure should interchange the isotropic subalgebras. Some obvious real forms are mentioned in Section 6; but we have not proven that these are all nor do we mean to conjecture that this is the case. We leave this for further study.

## ACKNOWLEDGEMENTS

It is a pleasure to thank Noureddine Mohammedi for his communications on the subject and for letting me know of his results [18] prior to their publication. I am grateful to Sonia Stanciu for useful conversations at the start of this project; and to Wafic Sabra for providing me with food (which really hit the spot!) during one particularly long computational session towards the end of the project. I would also like to thank Elias Kiritsis for his comments on a previous version of this paper and for making me aware of [17], which contains the first example of a solvable $N=2$ construction, albeit with $c=6$. Last but not least, I would like to thank the members of the String Theory group at QMW, and Chris Hull in particular, for their patience and their comments in the informal seminars I have given on this topic.